\documentclass[journal]{IEEEtran}
\ifCLASSINFOpdf
\else
\fi
\usepackage{graphicx}
\usepackage{amsmath,amssymb} % define this before the line numbering.
\usepackage{cite}
\usepackage{amsfonts}
\usepackage{algorithmic}
\usepackage{textcomp}
\usepackage{times}
\usepackage{epsfig}
\usepackage{mathrsfs}
\usepackage{multirow}
\usepackage{multicol}
\usepackage{subfigure}
\usepackage{url}
\usepackage{tabularx}
\usepackage{fdsymbol}

\newcolumntype{C}[1]{>{\centering\arraybackslash}p{#1}}
\newcommand{\etal}{\textit{et al}.}

\newcommand{\Rmnum}[1]{\expandafter\@slowromancap\romannumeral #1@}

\begin{document}

\title{Ultra High Fidelity Image Compression with $\ell_\infty$-constrained Encoding and Deep Decoding}

\author{Xi~Zhang and
        Xiaolin Wu,~\IEEEmembership{Fellow,~IEEE}% <-this % stops a space
\thanks{X.~Zhang is with the Department of Electronic Engineering, Shanghai Jiao Tong University, Shanghai, China (email: zhangxi\_19930818@sjtu.edu.cn).}
% <-this % stops a space
\thanks{X.~Wu is with the Department of Electrical \& Computer Engineering, McMaster University, Hamilton, L8G 4K1, Ontario, Canada (email: xwu@ece.mcmaster.ca).}}
% <-this % stops a space
% \thanks{Manuscript received April 19, 2005; revised August 26, 2015.}


\markboth{MANUSCRIPT SUBMITTED TO IEEE TRANSACTIONS ON IMAGE PROCESSING}%
{Shell \MakeLowercase{\textit{et al.}}: Bare Demo of IEEEtran.cls for IEEE Journals}



\maketitle

%GAN can remove large periodic errors of CALIC, image sharpness

\begin{abstract}
%Any deletion and addition of image features are absolutely forbidden in many computer vision applications, such as those

In many professional fields, such as medicine, remote sensing and sciences, users often demand image compression methods to be mathematically lossless.  But lossless image coding has a rather low compression ratio (around 2:1 for natural images).  The only known technique to achieve significant compression while meeting the stringent fidelity requirements is the methodology of $\ell_\infty$-constrained coding that was developed and standardized in nineties.  We make a major progress in $\ell_\infty$-constrained image coding after two decades, by developing a novel CNN-based soft $\ell_\infty$-constrained decoding method.  The new method repairs compression defects by using a restoration CNN called the $\ell_\infty\mbox{-SDNet}$ to map a conventionally decoded image to the latent image.  A unique strength of the $\ell_\infty\mbox{-SDNet}$ is its ability to enforce a tight error bound on a per pixel basis.  As such, no small distinctive structures of the original image can be dropped or distorted, even if they are statistical outliers that are otherwise sacrificed by mainstream CNN restoration methods.

%$\ell_\infty$-constrained decompression network enjoys the advantages of Convolutional Neural Networks (CNN) in image restoration and at the same time it also

More importantly, this research ushers in a new image compression system of $\ell_\infty$-constrained encoding and deep soft decoding ($\ell_\infty\mbox{-ED}^2$).
%, which couples the $\ell_\infty$-constrained predictive encoding and a CNN-based secondary decoding.
The $\ell_\infty \mbox{-ED}^2$ approach beats the best of existing lossy image compression methods (e.g., BPG, WebP, etc.) not only in $\ell_\infty$ but also in $\ell_2$ error metric and perceptual quality, for bit rates near the threshold of perceptually transparent reconstruction. Operationally, the new compression system is practical, with a low-complexity real-time encoder and a cascade decoder consisting of a fast initial decoder and an optional CNN soft decoder.

%Recently a number of CNN-based techniques were proposed to remove image compression artifacts. As in other restoration applications, these techniques all learn a mapping from decompressed patches to the original counterparts under the ubiquitous $\ell_2$ metric.
%However, this approach is incapable of restoring distinctive image details which may be statistical outliers but have high semantic importance (e.g., tiny lesions in medical images).  To overcome this weakness,

\end{abstract}


\IEEEpeerreviewmaketitle

\section{Introduction}

In many professional applications of computer vision, such as medicine, remote sensing, sciences and precision engineering, high spatial and spectral resolutions of images are always of paramount importance.  As the achievable resolutions of modern imaging technologies steadily increase, users are inundated by the resulting astronomical amount of image data. For example, a single pathology image generated by digital pathology slide scanner can easily reach the size of 1GB or larger.  For the sake of operability and cost-effectiveness, images have to be compressed for storage and communication in practical systems.

Unlike in consumer applications, such as smartphones and social media, where users are mostly interested in image esthetics, professionals of many technical fields are more concerned with the fidelity of decompressed images.  Ideally, they want mathematically lossless image compression, that is, the compression is an invertible coding scheme that can decode back to the original image, bit for bit identical.  Although the mathematically lossless image coding is the ultimate gold standard, its compression performance is too limited.  Despite years of research \cite{calic, lossless_calderbank, calic_TC, lossless_maniccam, lossless_boulgouris}, typical lossless compression ratios for medical and remote sensing images are only around 2:1, which fall far short of the requirements of most imaging and vision systems.

In order to meet the stringent fidelity requirements while still achieving significant compression ratio, the $\ell_\infty$-constrained (or colloquially called near-lossless) image coding methodology was developed \cite{near_ke, near_avcibas, near_chen, near_wu}
and standardized by ISO/JPEG \cite{calic,jpeg-ls}.  The distinction between the lossy and near-lossless compression methods is that the latter guarantees that at each pixel the absolute value of compression error is bounded by $\tau$, $\tau$ being a user specified error tolerance.  The tight per-sample error bound can only be realized by the minmax $\ell_\infty$ error criterion.  The ubiquitous $\ell_2$ error metric, which is adopted by consumer-grade lossy image compression methods, such as JPEG, JPEG 2000, WebP, etc., measures the average distortion over all pixels.  The $\ell_2$ compression is unable to preserve distinct image details that are statistical outliers but nevertheless vital to image semantics.  Such cases are common in machine vision applications; for examples, one is searching in a big ocean for a small boat, or a small lesion in a large organ.  When constrained by bit budget, an $\ell_2$-based lossy compression method tends to override such small structures by whatever dominant patterns in the background: ocean waves in the first example and liver textures in the second example.  In order to avert such risks
users (e.g., doctors, scientists and engineers) in many professions have to forego the $\ell_2$-based lossy compression widely used in consumer applications, and adopt the more conservative $\ell_\infty$ metric to keep compression error tolerance at a necessary minimum.

Since the standardization of the JPEG-LS nearlossless made in 1993, very little progress has been made in techniques for $\ell_\infty$ nearlossless image compression.  Scheuch et al.\ and Zhou et al.\ realized that the per-pixel $\ell_\infty$ error bound offers much stronger and useful information than in the $\ell_2$-based compression, and used it to mitigate compression noises in a process called soft decoding \cite{zhou2012}.  Soft decoding of the $\ell_\infty$-based nearlossless coded images is to solve the inverse problem of estimating the latent original image using the sparsity regularization and the prior knowledge of error bound $\tau$.  Although these methods are able to improve the precision of the compression reconstruction, their performances are limited by how well the assumed sparsity model fits the images in question.  Any further progress in soft decoding has to come from adopting a more versatile and precise statistical model for the inverse problem, whatever complex and defying analytical tools it might turn out to be.  The methodology of data-driven deep learning opens up such possibilities, as it can function as highly non-linear implicit statistical models.

Indeed, a large number of machine learning methods have been published recently for various image restoration tasks, including the reduction of compression noises \cite{ARCNN, CAR_galteri, CAR_guo}.  However, these deep learning based methods for soft decoding are apparently motivated by consumer and Internet applications and heavily influenced by mainstream lossy image/video compression methods.  They adopt ubiquitous $\ell_2$ or $\ell_1$ error metrics in the cost function of the restoration convolutional neural networks (CNN).  This average fidelity design criterion tends to smooth out rare distinct image features.  To counter the smoothing side effects, researchers widely adopt the technique of generative adversary neural network (GAN) for the task of compression artifacts removal.  With an emphasis on pleasing visual appearances rather than high objective fidelity, GAN has a well-known tendency to fabricate "realistic" looking but false details in the reconstructed images.  But any deletions and fabrications of image features are detrimental and should be absolutely forbidden in the professional fields of medicine, space, remote sensing, sciences, precision engineering and the alike.

% with a focus on perceptual image quality rather than high objective fidelity, and consequently not suited for critical professional applications targeted by this research.

One way to prevent the above identified side effects of minimizing the MSE loss and the GAN adversarial loss is the use of the $\ell_\infty$ loss function, when training the restoration CNN for compression artifact reduction or soft decoding.  Unlike MSE and GAN losses, the $\ell_\infty$ loss imposes a tight error bound on each single pixel; therefore, it can preserve distinct and subtle structures of the original image even if they are statistical outliers.
However, training the soft decoding CNN for minimum $\ell_\infty$ loss (called the $\ell_\infty$-SDNet hereafter) may have convergence difficulties, if the image compressor is designed for $\ell_2$ minimization as in current practice.  For the feasibility of the $\ell_\infty$-SDNet, we need to control the compression distortions at the source. The above reviewed $\ell_\infty$-constrained image coding approach serves our purpose perfectly.
The strict $\ell_\infty$ constraint of the encoder ensures each decoded pixel to have a given error bound $\tau$, and consequently offers the $\ell_\infty$-SDNet much needed strong priors to infer the inverse mapping of soft decoding.

The above reasoning leads to a novel ultra high-fidelity image compression system of $\ell_\infty$-constrained encoding and deep soft decoding ($\ell_\infty\mbox{-ED}^2$), which is the main contribution of this work.  In the $\ell_\infty\mbox{-ED}^2$ system, an image is $\ell_\infty$ coded by a sequential prediction-quantization encoder; the decoder is a cascade of the conventional (hard) decoder and a deep soft decoder that is the $\ell_\infty$-SDNet as outlined in the previous paragraph.  As the strict non-differentiable $\ell_\infty$ loss
does not permit backpropagation, we replace it with a differentiable quasi-$\ell_\infty$ loss term when optimizing the $\ell_\infty$-SDNet; also we modify the activation function of the network and make it respond to the $\ell_\infty$ criterion, expediting the quasi-$\ell_\infty$ training process.  In addition, we use the dilated convolution in the $\ell_\infty$-SDNet to achieve larger receptive field with fewer layers.  This makes
the $\ell_\infty$-SDNet shallower and more practical, while reducing risk of overfitting.

%vigilant against the side effects of the MSE loss and the adversarial loss.  The former criterion, due to the nature of $\ell_2$ metric, tends to smooth out subtle small image features; the latter criterion can fabricate false features.  These distortions, which are detrimental and should be prevented at all costs in the professional fields of medicine, space, engineering and sciences, can be suppressed by adding

The soft decoding network $\ell_\infty$-SDNet outperforms the best of existing lossy image compression methods such as BPG, WebP, J2K, not only in $\ell_\infty$ but also in $\ell_2$ error metric, for bit rates near the threshold of perceptually transparent reconstruction;
in effect, it reduces the critical bandwidth for perceptually lossless image compression.
The $\ell_\infty$ loss term contributes to the improved perceptual quality as $\ell_\infty$ penalizes the blurring of sharp edges more heavily than $\ell_2$.

%The $\ell_\infty$-CNN gives birth to a new hybrid image compression paradigm of coupling the $\ell_\infty$-constrained predictive encoding with a DCNN-based decoding.  We call it the strategy of light encoding and deep decoding (LEDD).

Finally, we stress the practical significance of this research.  The proposed $\ell_\infty\mbox{-ED}^2$ image compression system not only raises the bar for achievable rate-distortion performance in $\ell_\infty$, $\ell_2$ error metrics and in perceptual quality, but more importantly it can realize the compression gain in real time encoding.  This is simply because the $\ell_\infty\mbox{-ED}^2$ system uses the traditional low complexity predictive encoder.  Granted the CNN soft decoder $\ell_\infty$-SDNet is more expensive than conventional predictive decoder, but it is an optional refinement after the quick hard decoding.
The asymmetric complexity characteristic of the $\ell_\infty\mbox{-ED}^2$ system gives it a distinct operational advantage over the recently researched end-to-end pure CNN compression approach \cite{minnen2018,agustsson2018,li2018,mentzer2018}, as the former has a lower encoding complexity than the latter by order of magnitude.  Before a real-time CNN encoder of optimal rate-distortion performance can be economically implemented on end devices, such as cell phones, the $\ell_\infty\mbox{-ED}^2$ system allows practitioners to reap the benefits of deep learning in practical image compression systems.

%, having your cake and eating it too.

%It has practical significance for both professional and high-end consumer applications.

%complexity-scalable system

%better than POCS, interestingly better than pure l_2 optimization criterion

%This is the First and so far the Only paper on the subject.

%
%In building a deep neural network $G$ to map the $\ell_\infty$-encoded images to the corresponding latent images, we combine, like other authors of DCNN image restoration,
%the MSE loss, perceptual loss and adversarial loss.  However,

%Modern deep learning methodology opens up a new venue for ultra high-fidelity image compression. It demonstrates the potential of achieving the so-far best compression performance under the highly stringent minmax error metric.

The remainder of this paper is structured as follows.
Section \ref{sec:related} reviews the existing works for image compression artifacts removal.
Section \ref{sec:system} outlines the proposed $\ell_\infty\mbox{-ED}^2$ image compression system.
Section \ref{sec:sdnet} presents technical details of the $\ell_\infty$-constrained 
soft decoding network $\ell_\infty\mbox{-SDNet}$.
In Section \ref{sec:training}, details about the training of $\ell_\infty\mbox{-SDNet}$ are provided.
In section \ref{sec:performance}, we report our experimental results in comparison with other competing methods on four datasets.
Section \ref{sec:conclusion} concludes the paper.

%we conduct experiments to evaluate the effectiveness of the proposed $\ell_\infty$-constrained artifacts removal algorithm.

\begin{figure*}[!ht]
\centering
\includegraphics[width=\linewidth, height=0.22\linewidth]{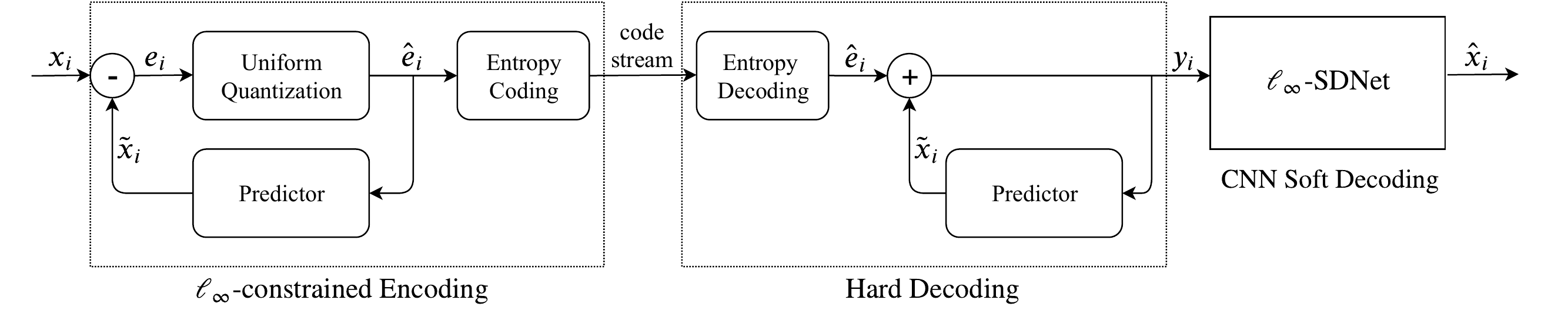}
\caption{Overall architecture of the $\ell_\infty\mbox{-ED}^2$ image compression system.
From left to right are the $\ell_\infty$-constrained encoding module, conventional hard decoding module and deep soft decoding ($\ell_\infty\mbox{-SDNet}$) module, respectively.}
\label{system}
\end{figure*}

\section{Related Works}
\label{sec:related}
%\subsection{Compression Artifacts Removal}

There is a rich body of literature on techniques for removing compression artifacts in images \cite{rw_foi, rw_zhang, rw_li, rw_chang, rw_dar, ARCNN, CAR_galteri, CAR_guo}.  The majority of the studies on the subject focus on postprocessing JPEG images to alleviate compression noises, apparently because JPEG is the most widely used lossy compression method.  The published works can be classified into two categories: explicit model-based methods and data-driven learning-based methods.

In the first category,
Reeve \etal \cite{rw_reeve} proposed to remove structured discontinuities of DCT code blocks by Gaussian filtering of the pixels around the DCT block boundaries.  This work was improved by
Zhai \etal \cite{rw_zhai} who performed postfiltering in shifted overlapped windows and fused the filtering results.  A total minimum variation method constrained by the JPEG quantization intervals was used by Alter \etal \cite{rw_alter} to reduce blocking artifacts and Gibbs phenomenon while preserving sharp edges.
Bredies \etal \cite{rw_bredies} studied optimality conditions of the TV minimization approach in infinite dimension, and used a primal-dual algorithm to solve a discrete version.
Li \etal \cite{rw_li} proposed to reduce compression artifacts by eliminating the artifacts that are part of the texture component, after decomposing images into structure and texture components.
Zhang \etal \cite{rw_zhang} approached the problem by merging two predictions of DCT coefficients in each block: one prediction is derived from nonlocal blocks of DCT coefficients and the other from quantized values of DCT coefficients.
Foi \etal \cite{rw_foi} proposed to use attenuated DCT coefficients to estimate the local image signal under an adaptive shape support.
Dar \etal \cite{rw_dar} formulated the compression post-processing procedure as a regularized inverse-problem for estimating the original signal given its reconstructed form.

In the class of data-driven learning-based methods, an early approach is sparse coding.
Chang \etal \cite{rw_chang} proposed to use a sparse dictionary learnt from a training image set to remove the block artifacts.
Liu \etal proposed a dual-dictionary method \cite{rw_liu} carried out jointly in the DCT and pixel domains.
Given the recent rapid development of deep convolutional neural networks (CNN), a number of CNN-based compression artifacts removal methods were published \cite{ARCNN, rw_svoboda, CAR_guo, CAR_galteri}.
Borrowing the CNN for super-resolution (SRCNN), Dong \etal \cite{ARCNN} proposed an artifact reduction CNN (ARCNN). The ARCNN has a three-layer structure: a feature extraction layer, a feature enhancement layer, and a reconstruction layer. This CNN structure is designed in the principle of sparse coding.  It was improved by Svoboda \etal \cite{rw_svoboda} who combined residual learning and symmetric weight initialization.
Zhang \etal \cite{dncnn} investigated the construction of feed-forward denoising convolutional neural networks (DnCNNs) in very deep architecture, learning algorithm, and regularization method into image denoising.
Recently, Guo \etal \cite{CAR_guo} and Galteri \etal \cite{CAR_galteri} proposed to reduce compression artifacts by Generative Adversarial Network (GAN), as GAN is able to generate sharper image details. It should be noted, however, that the GAN results may fabricate a lot of false hallucinated details, which is strictly forbidden in many scientific and medical applications.

To our best knowledge, the existing CNN-based image compression artifacts removal methods all focused on the JPEG post-processing, so we are the first to study the restoration of near-lossless compressed images.

\section{The $\ell_\infty\mbox{-ED}^2$ Image Compression System}
\label{sec:system}
In this section we present the design principle and details of the $\ell_\infty\mbox{-ED}^2$ image compression system.  The system architecture is graphically depicted in Fig.~\ref{system}.

First, we provide necessary background information on near-lossless image coding to facilitate the subsequent descriptions and understanding of our new work.
In the literature, near-lossless image coding refers to the $\ell_\infty$-constrained compression schemes that guarantee the compression error to be no larger than a user-specified bound for every pixel \cite{near_wu}.  This is typically realized within the framework of classic predictive coding as illustrated in Fig.~\ref{system}.  We only need to describe the encoder algorithm, as the decoder simply reverses the encoder.

Denoting by $X$ an image and $x_i$ the value of pixel $i$, image $X$ is compressed pixel by pixel sequentially, by first making a prediction of $x_i$:
\begin{align}
\tilde{x}_i = F(C_i)
\end{align}
where $C_i$ is a causal context that consists of previously coded pixels adjacent to $x_i$, and then entropy encoding and transmitting the prediction residual
\begin{align}
e_i = x_i - \tilde{x}_i.
\end{align}
At the decoder side, $x_i$ is recovered without any loss as $e_i + \tilde{x}_i$.  However, to gain higher compression ratio, one can quantize $e_i$ uniformly in step size $\tau$ to
\begin{align}
\hat{e}_i=
\begin{cases}
(2\tau + 1) \bigl\lfloor (e_i+\tau)/(2\tau+1) \bigr\rfloor \quad & e_i \geq 0 \\
(2\tau + 1) \bigl\lfloor (e_i-\tau)/(2\tau+1) \bigr\rfloor \quad & e_i < 0
\end{cases}
\label{quantization}
\end{align}
In this way, the decoded pixel value becomes $y_i = \hat{e}_i + \tilde{x}_i$, with quantization error
\begin{align}
\begin{split}
d &= x_i - y_i \\
  &= (e_i+\tilde{x}_i) - (\hat{e}_i + \tilde{x}_i) \\
  &= e_i - \hat{e}_i
\end{split}
\end{align}
But by Eq(\ref{quantization}), the quantization error will be no greater than the bound $\tau$ for every pixel:
\begin{align}
-\tau \leq x_i - y_i \leq \tau
\label{bound}
\end{align}
The above inequalities not only impose an $\ell_\infty$ error bound, but more importantly they, for the purpose of this work, provide highly effective priors, on per pixel, to optimize the deep neural networks for deep decoding of the $\ell_\infty$-compressed images.

\section{Design of $\ell_\infty \mbox{-SDNet}$}
\label{sec:sdnet}

\subsection{Overview}

Let $A$ and $A^{-1}$ be the encoder and decoder of an $\ell_\infty$-constrained near-lossless compression algorithm $A$, such as CALIC \cite{near_wu}, and let
$Y = A^{-1}A(X)$ be the conventional decoded image of coding the original image $X$.
The CNN restoration of the initially decoded image $Y$ can be considered as a soft decoding process, aiming to refine $Y$ and reconstruct an improved version $\hat{X}$ by maximally removing compression artifacts in $Y$.

To solve the problem of compression artifacts removal, we train the $\ell_\infty\mbox{-SDNet}$ (denoted by $G$) that takes decompressed image $Y$ as its input and generates the restored image $\hat{X} = G(Y)$.  In order to satisfy the stringent fidelity requirements of medical and scientific applications, the final output image $\hat{X}$ needs to be close to the original image $X$ not only perceptually but also mathematically.
To this end, we design our restoration network G with three new strategies.
First, we incorporate a so-called quasi-$\ell_\infty$ error bound into the cost function
%optimize the $\ell_\infty$-CNN with a new cost function
$L_G(X, G(Y))$ in optimizing the network:
\begin{align}
G = \arg \min_{G} \sum^{N}_{n=1} L_G(X_n, G(Y_n)),
\end{align}
for a given training dataset containing $N$ training samples $\{(X_n; Y_n)\}_{1 \leq n \leq N}$.  Second, we modify the activation function of the network and make it act on an $\ell_\infty$ criterion.  This second strategy is to reinforce the $\ell_\infty$ loss term of the first strategy.  The $\ell_\infty$ activation function has effects in both training and inference stages, while the quasi-$\ell_\infty$ loss function acts in the training only.
Third, we build the network $G$ by combining the encoder-decoder architecture and the dilated convolution, instead of the mainstream size-invariant fully convolutional network. This strategy is to achieve larger receptive field with fewer layers; the resulting shallower network has reduced risk of overfitting.
These three new designs will be detailed in the following three subsections; also
due to its $\ell_\infty$ properties we call the proposed network $\ell_\infty\mbox{-SDNet}$.

%Compared with existing works, the new cost function $L_G$, to be discussed in detail in the next section, adds an $\ell_\infty$ error bound in the reconstruction process.
%where $X_n$ is the $n$-th training sample in training set.

\begin{figure}[t]
\centering
\includegraphics[width=0.9\linewidth, height=0.7\linewidth]{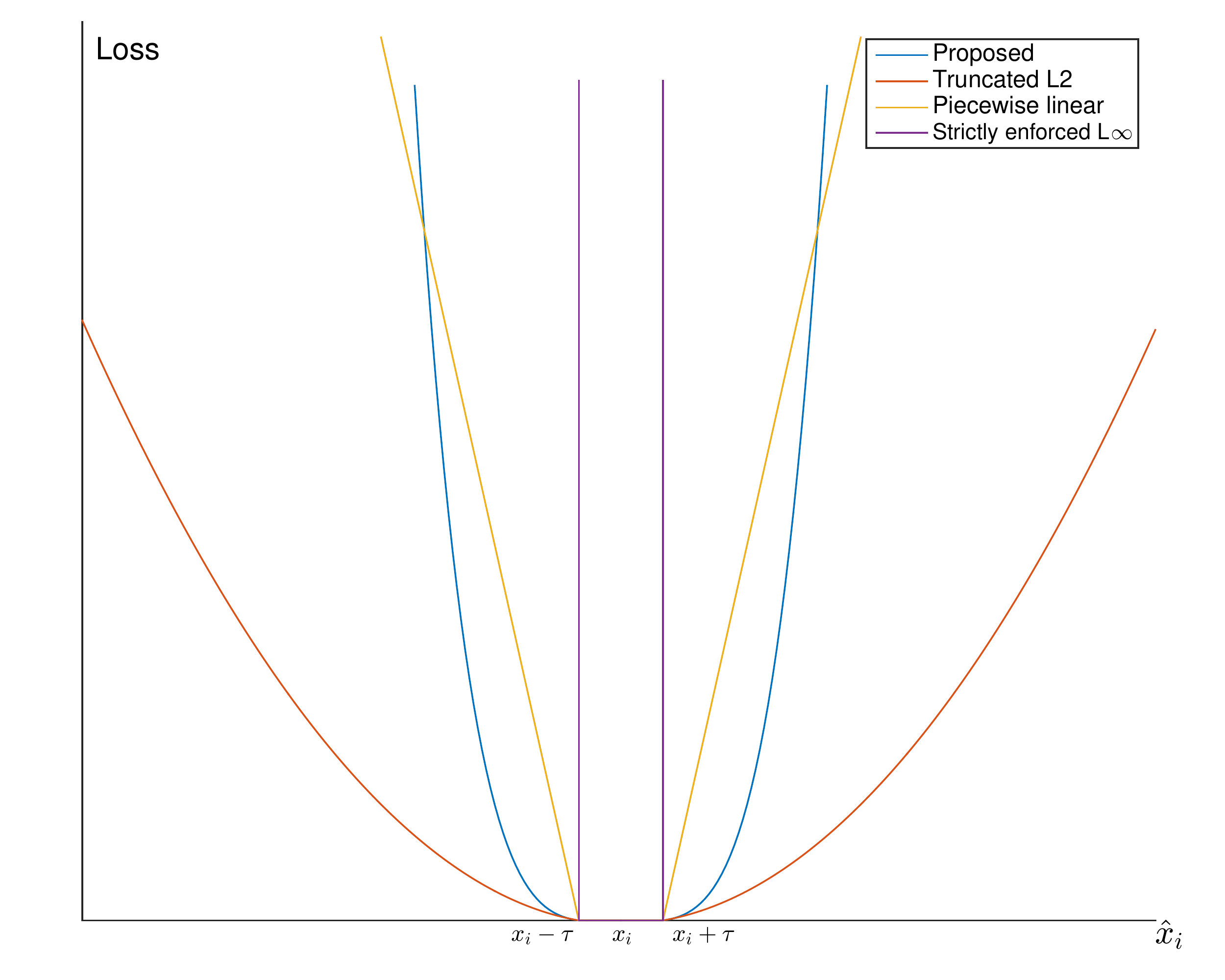}
\caption{True $\ell_\infty$ loss vs.\ its different approximations.
\textbf{Purple curve}: true $\ell_\infty$ loss; \textbf{yellow curve}: piecewise linear approximation; \textbf{red curve}: truncated $\ell_2$ loss; \textbf{blue curve}: the proposed quasi-$\ell_\infty$ loss.
}
\label{fig_infty}
\end{figure}

\subsection{Quasi-$\ell_\infty$ loss}
\label{infty_loss}
%In addition to the truncated activation function raised above,
%we also design an $\ell_\infty$-constrained loss function to further tighten the error bound on each pixel in the reconstructed image $\hat{X}$.

The existing CNN methods for compression artifacts removal adopt the ubiquitous $\ell_2$ loss function in optimizing the network:
\begin{align}
L_2 = \frac{1}{WH} \sum_{i} (\hat{x}_{i} - x_{i})^2
\end{align}
where $x_i$ and $\hat{x}_i$ are the values of pixel $i$ in $X$ and $\hat{X}$, $W$ and $H$ are the width and height of $X$.  Solely minimizing MSE seeks a good approximation in average sense, but it tends to smooth out distinctive image details which may be statistical outliers but have high semantic importance.  To prevent the smoothing artifacts and preserve high frequency structures, we incorporate the $\ell_\infty$ fidelity criterion of near-lossless compression into the optimization of the $\ell_\infty\mbox{-SDNet}$.
For each pixel $x_i$ in the restored image $\hat{X}$, the network objective function includes an $\ell_\infty$ loss term to heavily penalize the pixel values that are out of the range $[x_i-\tau, x_i+\tau]$, but not those within the range $[x_i-\tau, x_i+\tau]$.  A simple way is to modify the $\ell_2$ loss by making a flattened zero bottom portion of $[x_i-\tau, x_i+\tau]$, namely,
\begin{align}
L_- = \frac{1}{WH} \sum_{i} max \big( (\hat{x}_i - x_i)^2 - \tau^2, 0 \big)
\end{align}
However, this truncated $\ell_2$ loss is not steep enough outside the interval $[x_i-\tau, x_i+\tau]$ (see red curve in Fig.~\ref{infty_loss}); the penalties for pixel values beyond the user specified tolerance level are still too small.

If strictly enforced the $\ell_\infty$ loss should be infinity outside the interval $[x_i-\tau, x_i+\tau]$ (purple curve in Fig.~\ref{infty_loss}). However, when optimizing the network we cannot back-propagate the errors of the strict $\ell_\infty$ loss function since it is not differentiable outside the interval $[x_i-\tau, x_i+\tau]$.  The difficulty can be circumvented by modifying the $\ell_\infty$ loss function as a pair of steep slope lines arising from the ends of interval $[x_i-\tau, x_i+\tau]$ (yellow curve in Fig.~\ref{infty_loss}).  As the modified $\ell_\infty$ loss is a piecewise linear function, back propagation becomes possible.  Still some cautions are needed about the slope steepness; too steep a slop of the piecewise linear approximation of $\ell_\infty$ may penalize the values slightly exceeding the range $[x_i-\tau, x_i+\tau]$ too heavily, causing the training of the $\ell_\infty\mbox{-SDNet}$ to be unstable and difficult to converge.

To approximate the $\ell_\infty$ norm closely but without convergence difficulties, we propose a quasi-$\ell_\infty$ loss term:
\begin{align}
% L_\infty = -\frac{1}{WH} \sum_i log \big[ 1 -  max \big( |\hat{x}_i - x_i| - \tau, 0 \big) \big]
L_\infty = \frac{1}{WH} \sum_{i} max \big( (\hat{x}_i - x_i)^4 - \tau^4, 0 \big)
\end{align}
% where $\hat{x}_i$, $x_i$ and $\tau$ are normalized from $[0,255]$ to $[0,1]$ to ensure that the input of log function is positive.
As illustrated in Fig.~\ref{fig_infty}, the quasi-$\ell_\infty$ loss (blue curve) remains zero within the range $[x_i-\tau, x_i+\tau]$, but it rises once the pixel value drifts away from the range, and the rate of penalty increase is much higher  than the truncated $\ell_2$ loss (red curve).  When only slightly beyond the range $[x_i-\tau, x_i+\tau]$, the quasi-$\ell_\infty$ penalty is milder than
the piecewise linear version (yellow curve), and it jumps far more rapidly than the latter if the pixel value deviates any further.
Optimizing the $\ell_\infty\mbox{-SDNet}$ under the quasi-$\ell_\infty$ error metric forces the inference results to have a tight error bound on each pixel, while the training has a good convergence behaviour.

%\subsection{Joint optimization}
We combine the $\ell_2$ loss and the quasi-$\ell_\infty$ loss to optimize the proposed $\ell_\infty\mbox{-SDNet}$.
The joint loss function is defined as:
\begin{align}
  L_{G} = L_2 + \lambda L_{\infty}
\end{align}
where $\lambda$ is the hyper-parameter, which is set to $0.2$ in our experiments.

\begin{figure*}[!t]
\centering
\includegraphics[width=0.97\linewidth]{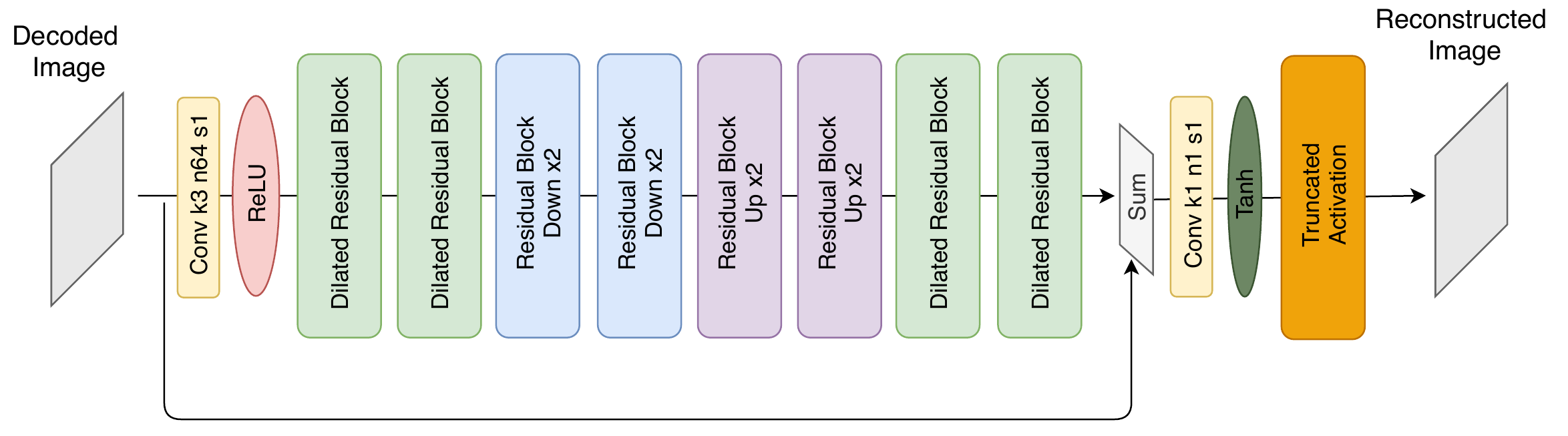}
\caption{Overall architecture of the proposed $\ell_\infty\mbox{-SDNet}$ with kernel size (k), number of feature maps (n) and stride (s).}
\label{netG}
\end{figure*}
\begin{figure*}[!t]
\centering
\includegraphics[angle=90, width=0.3\linewidth]{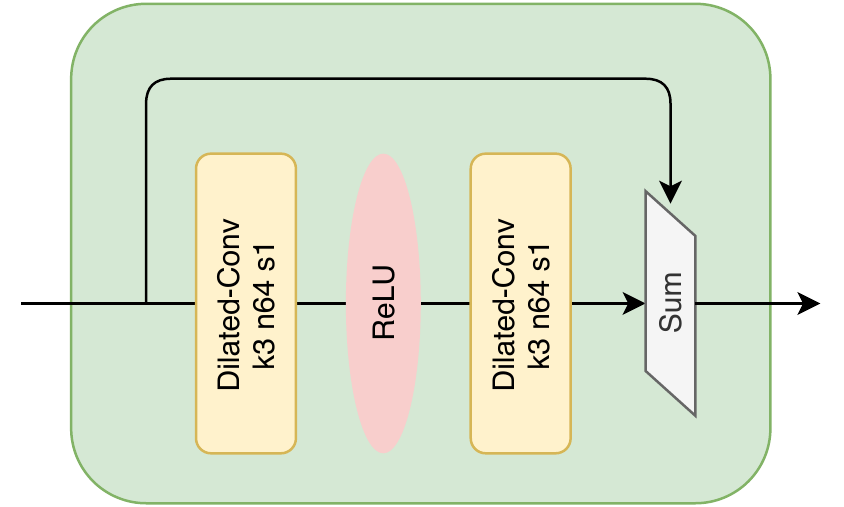}
\includegraphics[angle=90, width=0.3\linewidth]{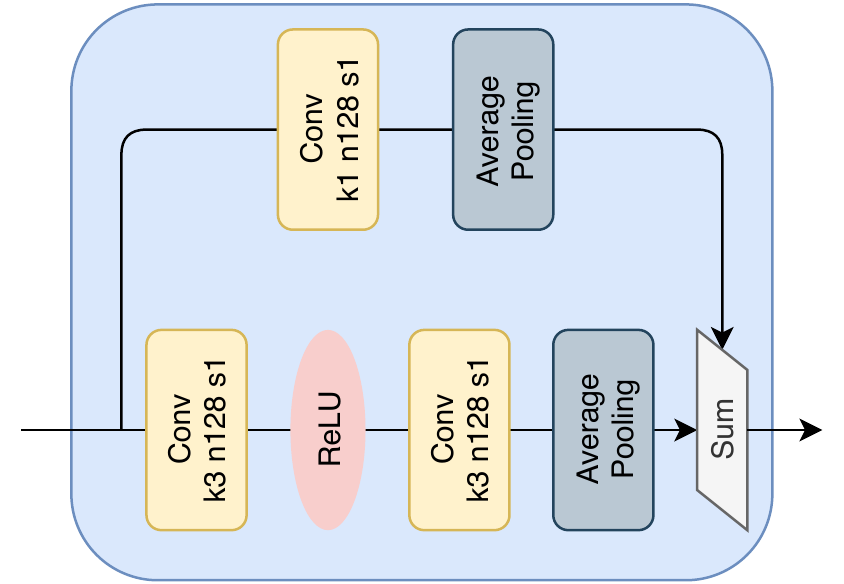}
\includegraphics[angle=90, width=0.3\linewidth]{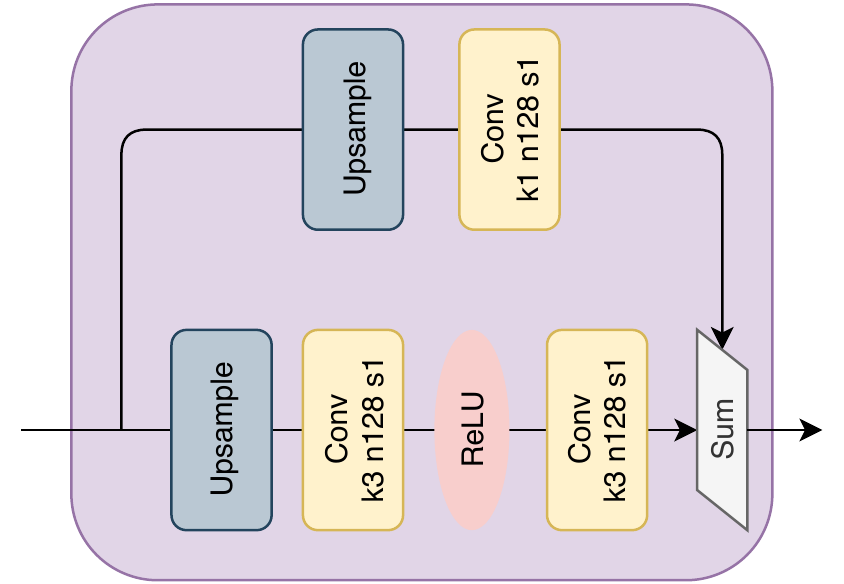}
\caption{Detailed architectures of the dilated residual block (left), downsampled residual block (middle) and upsampled residual block (right).}
\label{block}
\end{figure*}

\subsection{Truncated activation}
%why not true l_infty norm?

As the definition of the quasi-$\ell_\infty$ loss requires the knowledge of the ground truth value $x_i$, it can only be applied in the training stage not in the inference stage.  If the $\ell_\infty\mbox{-SDNet}$ operates on compressed images whose statistics does not match the distribution of the training images, it may not guarantee the $\ell_\infty$ tight error bound on every pixel.  In order to
strengthen the control of restoration errors of the $\ell_\infty\mbox{-SDNet}$ and improve its generalization ability, we design a novel activation function for the last layer of neurons in our network.

Given the error bound $\tau$ of near-lossless compression algorithm and the decompressed image $Y$, according to Eq(\ref{bound}), the reconstructed image $\hat{X}$ should satisfy the following constraint:
\begin{align}
y_i - \tau \leq \hat{x}_i \leq y_i + \tau
\label{x_i}
\end{align}
where index $i$ traverses all pixels in $\hat{X}$ and $Y$.
So, in order to ensure that $\hat{X}$ can satisfy such constraints in Eq.~\ref{x_i}, we need to clip the output $\tilde{X}$ of the $\ell_\infty\mbox{-SDNet}$ to generate $\hat{X}$, by
\begin{align}
\hat{x}_i = T(\tilde{x}_i) =
\begin{cases}
y_i-\tau, & \tilde{x}_i < y_i-\tau \\
\tilde{x}_i, & y_i-\tau \le \tilde{x}_i \le y_i+\tau \\
y_i+\tau, & \tilde{x}_i > y_i+\tau
\end{cases}
\label{activation}
\end{align}
where $\tilde{x}_i$ is the value of pixel $i$ in $\tilde{X}$.
The proposed truncation function can be implemented as a piecewise linear activation function embedded in the neural network. In the last layer of the $\ell_\infty\mbox{-SDNet}$, the activation function for each pixel is piecewise linear, similar to ReLU. The derivative of the truncated function is:
\begin{align}
T'(\tilde{x}_i) =
\begin{cases}
0, & \tilde{x}_i < y_i-\tau \\
1, & y_i-\tau \le \tilde{x}_i \le y_i+\tau \\
0, & \tilde{x}_i > y_i+\tau
\end{cases}
\label{derivative}
\end{align}

Note that $y_i$ is available from the decompressed image $Y$, so the truncation activation can be applied not only in the training phase, but also in the inference phase.  This truncated activation guarantees that the output of the $\ell_\infty\mbox{-SDNet}$ $\hat{x}_i$ falls into the interval $[y_i-\tau, y_i+\tau]$, and according to Eq.~\ref{bound}, $y_i$ is in $[x_i-\tau, x_i+\tau]$, so the reconstructed $\hat{x}_i$ must satisfy:
\begin{align}
-2\tau \leq \hat{x}_i - x_i  \leq 2\tau
\end{align}
The significance of the above analysis is that the restored image enjoys an error bound $2\tau$ on a per pixel basis even in the worst case.

\subsection{Network Architecture}
Recent works show that deeper and wider neural network architectures can achieve superior results in image restoration tasks~\cite{dncnn,dnrdb,symmetric,edsr,dbpn,rcan},
% WHY cite SR papers?
as they have sufficient capacity to learn a highly complex mapping.
Although removing compression noises is an image restoration problem, the compression noises are a well understood prior, and moreover in our case the distortion of the $\ell_\infty$-constrained compression is tightly bounded. As a result, the mapping from decoded images to lossless counterparts should be simpler than in general denoising and other restoration tasks.  For this reason the existing deep networks such as EDSR
%which have achieved excellent performance in image super-resolution task,
are not ideally suited to our task,
%the $\ell_\infty$-compressed image restoration task,
because too many convolution layers not only incur high computational costs but are also prone to over-fitting risk.

In the design of our $\ell_\infty\mbox{-SDNet}$, we try to make the network as simple as possible.  It contains 8 residual blocks \cite{ResNet}, not like 16 or 32 residual blocks used in existing image restoration CNNs.  However, reducing residual blocks narrows the receptive field, i.e., incapable to capture higher order statistical dependencies among pixels.
%In image restoration task, a pixel in the restored image only depends on a certain region of the input, which is called the receptive field.
%Intuitively, a larger region of the input can capture more context information. Therefore, the larger receptive field is desired for CNN so that no important features which are useful for restoring the current pixel are ignored.
This weakness is countered by two measures.  Firstly, we adopt an encoder-decoder architecture for the $\ell_\infty\mbox{-SDNet}$ with down-sampling and up-sampling operators, instead of a size-invariant fully convolutional architecture as in the mainstream design of image restoration networks.  Given the network depth, the former has a larger receptive filed than the latter due to the down-sampling operation.

Secondly, in each residual block, we replace the traditional convolutional layers with the dilated convolution layers~\cite{dilated,ddrnid,sirdc}.  By combining the encoder-decoder architecture and the dilated convolution, the proposed $\ell_\infty\mbox{-SDNet}$ can achieve the same or larger receptive field as other networks, but with fewer layers.
The overall architecture of the proposed $\ell_\infty\mbox{-SDNet}$ is illustrated in Fig.~\ref{netG} and the details of the residual block architecture are illustrated in Fig.~\ref{block}.  The dilation factor is set to $2$ in our design.

\section{Training of Multi-rate $\ell_\infty\mbox{-SDNet}$}
\label{sec:training}

In this section, we present the details of training the proposed  $\ell_\infty\mbox{-SDNet}$ for deep decoding of $\ell_\infty$-compressed images, including how to train the $\ell_\infty\mbox{-SDNet}$ for multiple compression bit rates so that it becomes universal and applicable on a wide range of bit rates (compression ratios).

%compression artifacts removal.
%conduct experiments to evaluate the effectiveness of the proposed $\ell_\infty$-constrained artifacts removal algorithm for  near-lossless decompression.
%All experiments are implemented with a NVIDIA Titan X GPU.

\subsection{Training data and settings}
In the existing works on CNN-based compression artifacts removal \cite{CAR_guo,CAR_galteri,ARCNN}, data used for training are from the popular datasets like BSD100, ImageNet or MSCOCO.
But images in these datasets are already compressed and have relatively low resolutions, hence they are not suitable as the ground truth for our purpose of ultra high fidelity image decompression for professional applications. Instead, we choose the high-quality uncompressed image dataset DIV2K for training.  The DIV2K dataset consists of 2K resolution images and is commonly used to synthesize paired training images for the construction of image restoration CNNs.  To generate compressed and original image pairs to train the above proposed $\ell_\infty\mbox{-SDNet}$, we use the $\ell_\infty$-constrained (near-lossless) CALIC algorithm \cite{near_wu} to compress the DIV2K images with a given error bound $\tau$ for every pixel ($\tau=1,2,3,4,5,6,7,8$ used in our experiments).

The DIV2K high resolution images are of consumer RGB type.  We carry out our experiments on the Y (luminance) channel of the DIV2K images because the Y channel has the most information (highest entropy) and hence most difficult to compress.  Here we deliberately challenge ourselves by using high resolution consumer monochorme images to train the $\ell_\infty\mbox{-SDNet}$, but carrying out inferences beyond the DIV2K dataset, including some satellite images.

%Typically, compression algorithms work in the YUV color space, to separate luminance from chrominance information. For this reason, we convert all images into YUV color space and conduct experiments on the luminance channel.

%\subsection{Training Details}
%
%DIV2K dataset
%
%ours --> l_\infty-cnn
%
%Typical compression algorithms work in the YUV color space to separate luminance from chrominance information, and sub-sample chrominance, since the human visual system is less sensitive to its changes. For this reason, in our experiments, we convert all images to YUV space and then just conduct experiments on Y channel.

The training images are decomposed into $128 \times 128$ sub-images with stride $32$, after compressed by the near-lossless CALIC algorithm.
We train the proposed $\ell_\infty\mbox{-SDNet}$ with Adam optimizer~\cite{adam} by setting momentum term $\beta_1=0.9$ and $\beta_2=0.999$.
The neural network is trained with $100$ epochs at the learning rate of $10^{-4}$ and other $50$ epochs with learning rate of $10^{-5}$.
We implement the proposed model in TensorFlow~\cite{TF} and train it with 4 NVIDIA TITAN Xp GPUs.

\subsection{Multi-rate Training}
In previous CNN-based methods for compression artifacts removal, the training is carried out only with respect to a single compression bit rate.  For instance, the training images are all of the same quality factor (QF) of the JPEG compression standard.  Needless to say, the CNN learned for a given QF may not work well for images compressed in different QF's.  Alternatively, many QF-specific CNNs can be trained for different QF's, but this approach is inefficient in practice.  In this work, we train a unified $\ell_\infty\mbox{-SDNet}$ for restoring images compressed in a range of bit rates.  Each image in the training set is compressed for different, from low to high compression ratios, or for increasing $\ell_\infty$ bounds $\tau = 1, \cdots, 8 $.  Thus, each patch in the original image has multiple compressed versions of different qualities, forming multiple sample pairs; all of them participate in the training of the multi-rate $\ell_\infty\mbox{-SDNet}$.  It turns out, as shown in the ablation study section, that the resulting multi-rate $\ell_\infty\mbox{-SDNet}$ is more robust than the single-rate counterpart.  The multi-rate $\ell_\infty\mbox{-SDNet}$ even outperforms the single-rate $\ell_\infty\mbox{-SDNet}$ on testing images that are compressed at the said single rate.

%Besides, for one training sample, in single-rate training, neural network is optimized to learn a mapping from the given-rate encoded patch to the corresponding original patch, but in multi-rate training, neural network is optimized to learn a mapping from many different-rate encoded patches to the same original patch. This many-to-one training process contributes to learning features from images compressed at different rates.
%Our experimental results demonstrate that multi-rate training can achieve higher PSNR and SSIM than single-rate training in compression artifacts removal task.

\section{Performance Evaluation}
\label{sec:performance}

We have implemented the proposed $\ell_\infty\mbox{-SDNet}$ for the task of removing compression artifacts, and conducted extensive experiments of near-lossless image decompression with it.
Recall that our objective is to achieve the best possible compression rate-distortion performance at the threshold of perceptually lossless quality by coupling the CNN decoding and the $\ell_\infty$-constrained predictive encoding of images.  To establish our claim, we compare our results with those of three popular lossy image compression methods: JPEG 2000~\cite{jpeg2000}, WebP~\cite{webp} and BPG~\cite{bpg} in both $\ell_\infty$ and $\ell_2$ distortion metrics when they operate at perceptually lossless quality level.

JPEG 2000 is an image compression standard created by the Joint Photographic Experts Group committee in 2000 with the intention of superseding the original JPEG standard.  WebP is an image format developed by Google, announced in 2010 as a new open standard for image compression.  BPG is the state-of-the-art image lossy compression method of the best performance so far.  It is based on the intra-frame encoding of the High Efficiency Video Coding (HEVC) video compression standard.  Also, we include in our comparison group the deep learning method DnCNN that can be applied for compression artifact removal \cite{dncnn}.  The DnCNN is trained with the same DIV2K and the near-lossless CALIC-compressed image pairs, as in the training of the proposed $\ell_\infty\mbox{-SDNet}$.

%not an end-to-end pure CNN approach,
% the second motivation, which is perhaps even more important in the perspective of general perceptually transparent lossy compression,

%We also compare the proposed $\ell_\infty$-CNN with CNN-based method DnCNN~\cite{dncnn} after retraining the latter with our datasets.  Recently, some new methods are proposed to utilize the transform domain information to achieve better recovery results for the task of image compression artifacts removal~\cite{dual,CAR_guo}.  However, we cannot compare with these methods, as they are only suitable to the transform coding methods but the near-lossless compression is implemented in the predictive coding framework.
% considering that the near-lossless image compression is realized in the predictive coding framework.
% Therefore, we cannot compare with these methods.
% image compression artifacts removal, which

\begin{figure}[!t]
\centering
\includegraphics[height=0.32\linewidth]{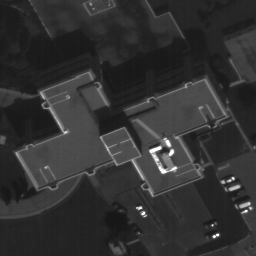}
\includegraphics[height=0.32\linewidth]{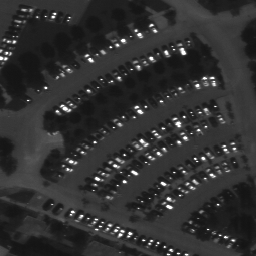}
\includegraphics[height=0.32\linewidth]{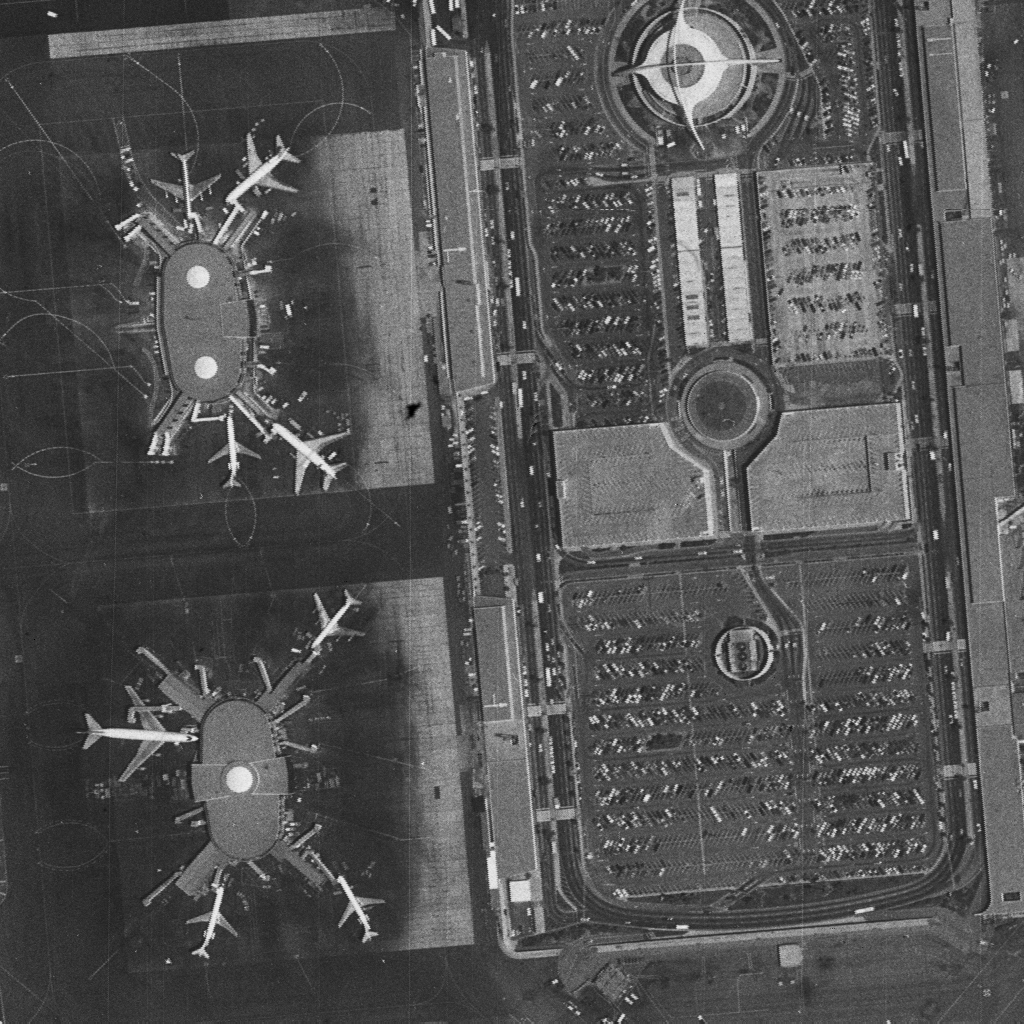}
\caption{Sample images of the Aerial dataset.}
\label{aerial22}
\end{figure}

Although the CNN decompression methods are trained by the DIV2K dataset, they and the other non-CNN methods are tested on four different test image sets LIVE1~\cite{LIVE1}, Kodak~\cite{Kodak} and Urban~\cite{Urban100} and a set of  aerial and satellite image.   The first three test image sets are widely used in literature for evaluating image restoration methods.
The last test set has 22 high-resolution aerial and satellite images, which are typical of those in remote sensing and other professional applications.
Some sample images from the set are shown in Fig.~\ref{aerial22}.
\begin{figure}[!t]
\centering
\vskip -0.4cm
\includegraphics[width=\linewidth, height=0.8\linewidth]{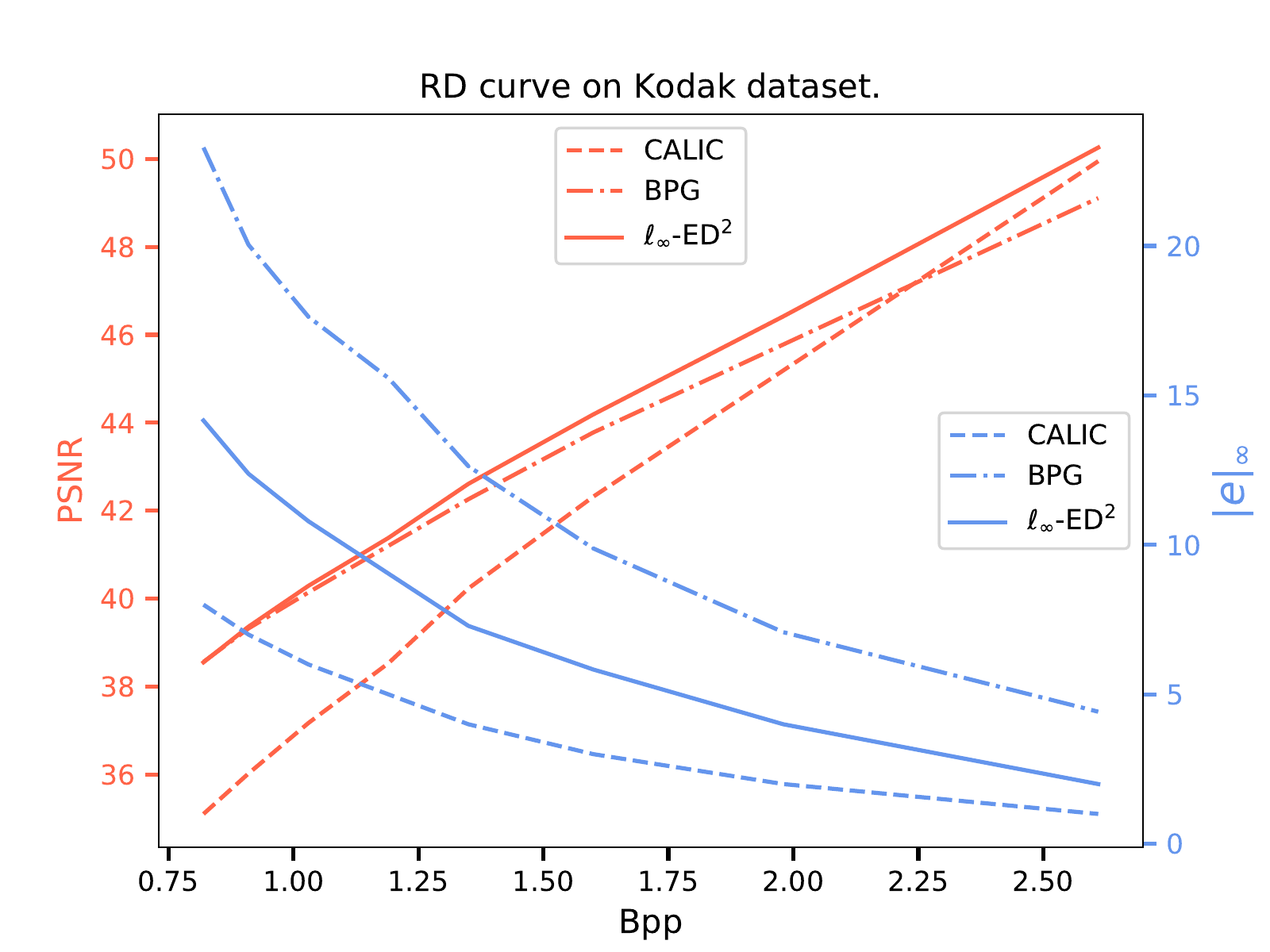}
\caption{PSNR and $\|e\|_\infty$ of the competing methods on the the Kodak dataset.}
\label{psnr_infty}
\end{figure}

\subsection{Quantitative evaluation}

To facilitate fair rate-distortion performance evaluations, for each test image, the rates of JPEG2000, WebP and BPG are adjusted to match that of the near-lossless CALIC.  Given the rate, both $\ell_\infty$ and $\ell_2$ (PSNR) distortion metrics are used to measure the image quality of different methods.
The $\ell_\infty$ error is defined as follow:
\begin{align}
\centering
\|e\|_\infty = max \Big( |\hat{x}_i - x_i| \Big)
\end{align}

Rate-distortion performance results of the competing methods are tabulated in Tables~\ref{tab_live1}, \ref{tab_kodak}, \ref{tab_urban}, \ref{tab_aerial}, for the four test sets respectively.
% RD curves of the competing methods are shown in Fig.~\ref{rd_curves}.
Note the reported value of $\|e\|_\infty$ is the average for the corresponding test set.
As shown in the four tables, the proposed $\ell_\infty\mbox{-ED}^2$ system outperforms all lossy compression methods JPEG2000, WebP and BPG consistently, not only in PSNR but also in $\|e\|_\infty$, for bit rate larger than 0.8 bpp (roughly the threshold of perceptually transparent reconstruction).  At the same time, the proposed $\ell_\infty\mbox{-SDNet}$ also beats the CNN-based denoising model DnCNN in both PSNR and $\|e\|_\infty$. It is noteworthy that the former has the smaller model size and faster inference speed than the latter.  Table~\ref{tab_size} shows that the $\ell_\infty\mbox{-SDNet}$ is only half the size of DnCNN and its inference speed is three times faster.
\begin{table}[!t]
\caption{Model size and inference time for a 1024$\times$1024 image in GPU.}
\begin{center}
\renewcommand\arraystretch{1.5}
\begin{tabular}{c|C{2cm}C{2cm}}
\hline
  & DnCNN\cite{dncnn} & $\ell_\infty\mbox{-SDNet}$ \\
\hline
% \hline
Model size (MB) & 2.24 & 1.08 \\
\hline
Inference time (s) & 0.22 & 0.08 \\
\hline
\end{tabular}
\end{center}
\label{tab_size}
\end{table}
\begin{table*}[!ht]
\caption{Performance results (PSNR/$\|e\|_\infty$) of the competing methods on the LIVE1 dataset.}
\begin{center}
\renewcommand\arraystretch{1.15}
\begin{tabular}{C{2cm}|C{2cm}C{2cm}C{2cm}C{2cm}C{2cm}C{2cm}}
\hline
Bit rate (bpp) & CALIC & JPEG2000 & WebP & BPG & DnCNN & $\ell_\infty\mbox{-ED}^2$ \\
 % & & & & & & \\ 
\hline
\hline
2.78 & 49.93\ /\ 1.00 & 47.76\ /\ 5.51 & 45.73\ /\ 6.31 & 48.99\ /\ 4.38 & 50.14\ /\ 2.00 & \textbf{50.23\ /\ 2.00} \\

2.15 & 45.19\ /\ 2.00 & 44.36\ /\ 8.72 & 43.39\ /\ 9.38 & 45.42\ /\ 7.24 & 46.12\ /\ 4.00 & \textbf{46.28\ /\ 4.00} \\

1.76 & 42.31\ /\ 3.00 & 42.35\ /\ 11.76 & 41.41\ /\ 12.34 & 43.26\ /\ 10.34 & 43.68\ /\ 6.00 & \textbf{43.92\ /\ 5.82} \\

1.50 & 40.20\ /\ 4.00 & 40.64\ /\ 14.51 & 40.00\ /\ 14.89 & 41.72\ /\ 12.86 & 42.02\ /\ 7.82 & \textbf{42.25\ /\ 7.20} \\

1.31 & 38.49\ /\ 5.00 & 39.41\ /\ 16.72 & 38.83\ /\ 17.80 & 40.49\ /\ 15.76 & 40.63\ /\ 9.79 & \textbf{40.91\ /\ 9.03} \\

1.15 & 37.13\ /\ 6.00 & 38.32\ /\ 20.66 & 38.11\ /\ 19.10 & 39.47\ /\ 19.24 & 39.49\ /\ 11.88 & \textbf{39.79\ /\ 10.86} \\

1.03 & 35.98\ /\ 7.00 & 37.39\ /\ 24.93 & 36.92\ /\ 24.41 & 38.65\ /\ 20.90 & 38.45\ /\ 13.82 & \textbf{38.81\ /\ 12.20} \\

0.94 & 35.05\ /\ 8.00 & 36.62\ /\ 26.34 & 36.55\ /\ 24.28 & 37.91\ /\ 23.93 & 37.66\ /\ 15.76 & \textbf{38.05\ /\ 14.01} \\

\hline
\end{tabular}
\end{center}

\label{tab_live1}
\end{table*}

\begin{table*}[!ht]
\caption{Performance results (PSNR/$\|e\|_\infty$) of the competing methods on the Kodak dataset.}
\begin{center}
\renewcommand\arraystretch{1.15}
\begin{tabular}{C{2cm}|C{2cm}C{2cm}C{2cm}C{2cm}C{2cm}C{2cm}}
\hline
Bit rate (bpp) & CALIC & JPEG2000 & WebP & BPG & DnCNN & $\ell_\infty\mbox{-ED}^2$ \\
 % & & & & & & \\ 
\hline
\hline
2.61 & 49.95\ /\ 1.00 & 47.96\ /\ 5.21 & 45.86\ /\ 6.33 & 49.11\ /\ 4.42 & 50.15\ /\ 2.00 & \textbf{50.26\ /\ 2.00} \\

1.98 & 45.19\ /\ 2.00 & 44.76\ /\ 8.42 & 43.74\ /\ 8.92 & 45.78\ /\ 7.08 & 46.25\ /\ 4.00 & \textbf{46.42\ /\ 4.00} \\

1.60 & 42.32\ /\ 3.00 & 42.78\ /\ 11.58 & 41.91\ /\ 11.80 & 43.78\ /\ 9.88 & 43.97\ /\ 6.00 & \textbf{44.19\ /\ 5.83} \\

1.35 & 40.23\ /\ 4.00 & 41.20\ /\ 14.12 & 40.55\ /\ 14.42 & 42.26\ /\ 12.63 & 42.35\ /\ 8.00 & \textbf{42.61\ /\ 7.29} \\

1.19 & 38.53\ /\ 5.00 & 40.07\ /\ 15.80 & 39.53\ /\ 16.33 & 41.20\ /\ 15.58 & 41.11\ /\ 9.92 & \textbf{41.38\ /\ 9.04} \\

1.03 & 37.17\ /\ 6.00 & 38.98\ /\ 19.46 & 38.53\ /\ 19.54 & 40.14\ /\ 17.63 & 39.96\ /\ 11.84 & \textbf{40.29\ /\ 10.79} \\

0.91 & 36.02\ /\ 7.00 & 38.11\ /\ 24.38 & 37.67\ /\ 22.50 & 39.32\ /\ 20.04 & 38.98\ /\ 13.79 & \textbf{39.36\ /\ 12.38} \\

0.82 & 35.10\ /\ 8.00 & 37.31\ /\ 27.10 & 36.99\ /\ 24.41 & \textbf{38.56}\ /\ 23.29 & 38.12\ /\ 15.80 & 38.55\ /\ \textbf{14.16} \\

\hline
\end{tabular}
\end{center}

\label{tab_kodak}
\end{table*}

\begin{table*}[!ht]
\caption{Performance results (PSNR/$\|e\|_\infty$) of the competing methods on the Urban100 dataset.}
\begin{center}
\renewcommand\arraystretch{1.15}
\begin{tabular}{C{2cm}|C{2cm}C{2cm}C{2cm}C{2cm}C{2cm}C{2cm}}
\hline
Bit rate (bpp) & CALIC & JPEG2000 & WebP & BPG & DnCNN & $\ell_\infty\mbox{-ED}^2$ \\
 % & & & & & & \\ 
\hline
\hline
2.66 & 49.97\ /\ 1.00 & 47.74\ /\ 5.76 & 45.94\ /\ 6.92 & 49.34\ /\ 4.20 & 50.29\ /\ 2.00 & \textbf{50.42\ /\ 2.00} \\

2.06 & 45.30\ /\ 2.00 & 44.18\ /\ 9.51 & 43.23\ /\ 10.35 & 45.33\ /\ 7.10 & 46.32\ /\ 4.00 & \textbf{46.47\ /\ 4.00} \\

1.71 & 42.38\ /\ 3.00 & 42.00\ /\ 13.43 & 41.28\ /\ 13.38 & 43.22\ /\ 9.71 & 43.93\ /\ 6.00 & \textbf{44.12\ /\ 5.93} \\

1.45 & 40.24\ /\ 4.00 & 40.15\ /\ 16.21 & 39.72\ /\ 16.76 & 41.82\ /\ 13.16 & 42.19\ /\ 7.95 & \textbf{42.44\ /\ 7.57} \\

1.28 & 38.54\ /\ 5.00 & 38.84\ /\ 20.39 & 38.44\ /\ 20.24 & 40.73\ /\ 16.11 & 40.86\ /\ 9.80 & \textbf{41.14\ /\ 9.20} \\

1.13 & 37.17\ /\ 6.00 & 37.67\ /\ 24.50 & 37.35\ /\ 23.84 & 39.75\ /\ 19.58 & 39.74\ /\ 11.75 & \textbf{40.02\ /\ 11.02} \\

1.02 & 36.01\ /\ 7.00 & 36.69\ /\ 28.54 & 36.46\ /\ 27.14 & 38.82\ /\ 22.80 & 38.61\ /\ 13.88 & \textbf{39.04\ /\ 12.70} \\

0.94 & 34.96\ /\ 8.00 & 35.92\ /\ 31.64 & 35.84\ /\ 29.63 & 38.12\ /\ 25.91 & 37.86\ /\ 15.82 & \textbf{38.32\ /\ 14.38} \\

\hline
\end{tabular}
\end{center}

\label{tab_urban}
\end{table*}

\begin{table*}[!ht]
\caption{Performance results (PSNR/$\|e\|_\infty$) of the competing methods on the Aerial dataset.}
\begin{center}
\renewcommand\arraystretch{1.15}
\begin{tabular}{C{2cm}|C{2cm}C{2cm}C{2cm}C{2cm}C{2cm}C{2cm}}
\hline
Bit rate (bpp) & CALIC & JPEG2000 & WebP & BPG & DnCNN & $\ell_\infty\mbox{-ED}^2$ \\
 % & & & & & & \\ 
\hline
\hline
2.45 & 49.91\ /\ 1.00 & 47.54\ /\ 5.41 & 46.05\ /\ 5.90 & 48.72\ /\ 4.27 & 50.10\ /\ 2.00 & \textbf{50.27\ /\ 2.00} \\

1.80 & 45.25\ /\ 2.00 & 44.04\ /\ 8.02 & 43.38\ /\ 8.95 & 45.18\ /\ 7.10 & 46.15\ /\ 4.00 & \textbf{46.35\ /\ 3.77} \\

1.44 & 42.39\ /\ 3.00 & 42.00\ /\ 11.50 & 41.51\ /\ 11.45 & 43.15\ /\ 9.23 & 43.80\ /\ 5.92 & \textbf{43.98\ /\ 5.36} \\

1.20 & 40.30\ /\ 4.00 & 40.42\ /\ 13.77 & 40.02\ /\ 14.68 & 41.54\ /\ 12.41 & 41.89\ /\ 7.80 & \textbf{42.17\ /\ 7.13} \\

1.02 & 38.64\ /\ 5.00 & 39.27\ /\ 16.63 & 38.97\ /\ 16.50 & 40.50\ /\ 14.63 & 40.65\ /\ 9.76 & \textbf{40.92\ /\ 8.82} \\

0.87 & 37.35\ /\ 6.00 & 38.35\ /\ 19.95 & 37.86\ /\ 19.95 & 39.28\ /\ 17.45 & 39.51\ /\ 11.78 & \textbf{39.84\ /\ 10.59} \\

0.77 & 36.18\ /\ 7.00 & 37.56\ /\ 23.27 & 37.15\ /\ 22.13 & 38.53\ /\ 21.02 & 38.41\ /\ 13.60 & \textbf{38.84\ /\ 12.23} \\

0.69 & 35.15\ /\ 8.00 & 36.91\ /\ 26.81 & 36.51\ /\ 24.72 & 37.91\ /\ 22.68 & 37.66\ /\ 15.62 & \textbf{38.16\ /\ 13.95} \\

\hline
\end{tabular}
\end{center}

\label{tab_aerial}
\end{table*}

%Note that the maximum error magnitudes $\|e\|_\infty$ should be an integer between $[0,255]$, but we show the $\|e\|_\infty$ results averaged over a dataset, which turn out to be the floating point numbers.

% \begin{figure*}[!ht]
% \centering
% \includegraphics[width=0.45\linewidth, height=0.345\linewidth]{figure/rd_live1}
% \includegraphics[width=0.45\linewidth, height=0.345\linewidth]{figure/rd_kodak}\\
% \includegraphics[width=0.45\linewidth, height=0.345\linewidth]{figure/rd_urban}
% \includegraphics[width=0.45\linewidth, height=0.345\linewidth]{figure/rd_aerial}
% \caption{Performance results (RD curves) of the competing methods on the LIVE1, Kodak, Urban100 and Aerial22 datasets.}
% \label{rd_curves}
% \end{figure*}

Fig.~\ref{psnr_infty} plots the PSNR (left Y axis) and $\|e\|_\infty$ (right Y axis) of different methods vs.\ bit rate for the Kodak dataset (the patterns are similar for other datasets).  The near-lossless CALIC is the best in the $\ell_\infty$ norm, but the worst in the $\ell_2$ norm (PSNR). The proposed deep decoding technique greatly improves the PSNR of CALIC while still maintaining a relatively tight $\ell_\infty$ error bound on each pixel.  More importantly, the $\ell_\infty\mbox{-ED}^2$ system method achieves higher PSNR than the BPG method for bit rates near and above the threshold of perceptually transparent reconstruction (bpp $>$ 0.76, PSNR $<$ 38); at the same time its $\|e\|_\infty$ is only half of the latter.  This is quite remarkable considering BPG is optimized for $\ell_2$ and it is considered the best PSNR performer up to now.

\begin{figure}[!t]
\centering
\includegraphics[width=\linewidth, height=0.8\linewidth]{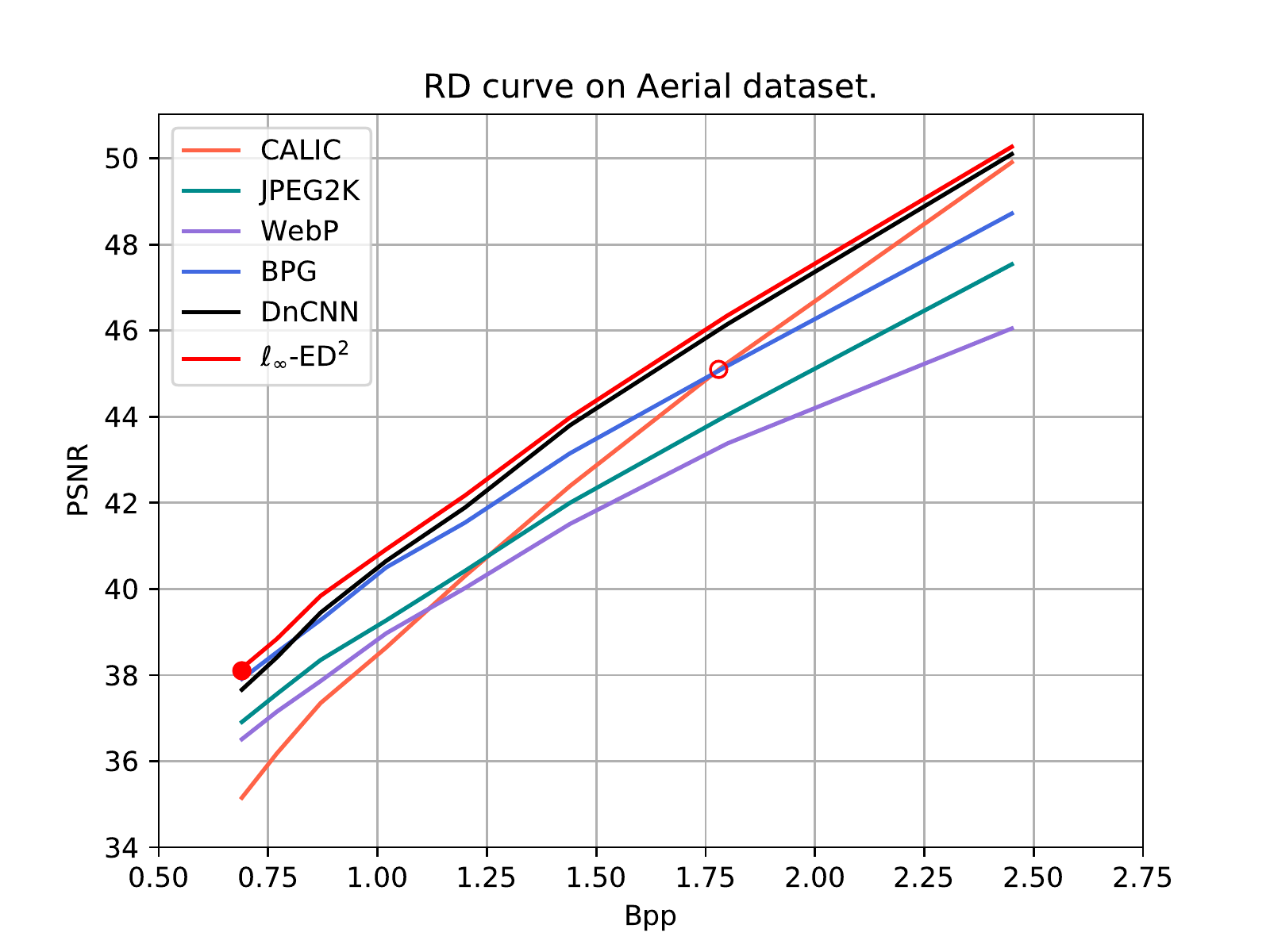}
\caption{Rate-distortion curves of the competing methods on the Aerial dataset.}
\label{rd_curves}
\end{figure}

Fig.~\ref{rd_curves} graphically represent the rate-distortion behaviours of different methods for the Aerial dataset (the patterns are similar for other datasets).  In terms of compression methodology, JPEG 2000, WebP and BPG are transform based coding, while CALIC is predictive coding.  Before this work the consensus is that predictive coding outperforms transform coding only for very high bit rates (low compression ratio), hence it is suited for lossless but not for lossy compression.  The above long held view has been now changed by the $\ell_\infty\mbox{-ED}^2$ system.  As shown in Fig.~\ref{rd_curves}, the CNN soft decoding of predictive coded images outperforms all transform coding methods, including the state-of-the-art BPG method.  Without the CNN soft decoding, the rate-distortion curves of the transform coding method BPG and the predictive coding method CALIC cross each other at 1.76 bpp.  With the CNN soft decoding, the cross over point moves to a much lower bit rate of 0.73 bpp.  These findings suggest that the best practical strategy for perceptually transparent lossy image compression is the $\ell_\infty$-constrained predictive encoding followed by the CNN soft decoding.
This strategy allows us to use a low complexity encoder for real time image acquisition and a
fast preview decoder, while still having the option to achieve the original quality with a tight per pixel error bound.

%instead of end-to-end CNN compression
%progressive decoding, secondary refinement

% perform predictive coding at high rate, so that we can achieve real-time encoding and control distortions to some modest level, As a result, CNN secondary decoding can successfully repair the distortions

% this research ushers in a new hybrid image
% compression strategy, called light encoding and deep decoding
% (LEDD), which couples the $\ell_\infty$-constrained predictive encoding
% and a DCNN-based decoding. The LEDD strategy beats or
% matches the best of existing lossy image compression methods
% such as BPG, WebP, J2K, not only in $\ell_\infty$ but also in $\ell_2$ error
% metric and perceptual quality, for bit rates near the threshold of
% perceptually transparent reconstruction. Moreover, The LEDD
% has a very low encoding complexity and hence is suited for realtime
% applications on end devices.

% \subsection{Encoding complexity}
% \input{tabs/encoding_time}

\begin{figure*}[!ht]
\centering
\vskip -0.4cm
\includegraphics[width=\linewidth]{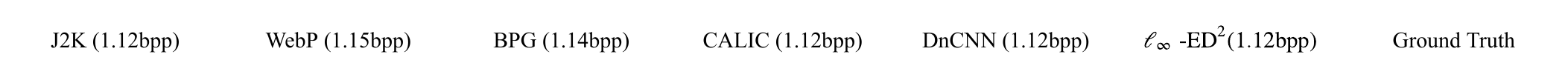}
\vskip -0.2cm
\includegraphics[width=\linewidth, height=0.14\linewidth]{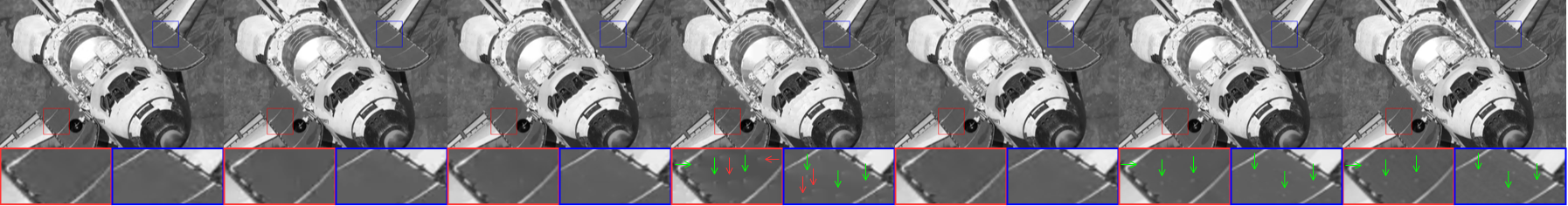}
\includegraphics[clip, trim=0.8cm 0cm 0.4cm 0cm, width=\linewidth]{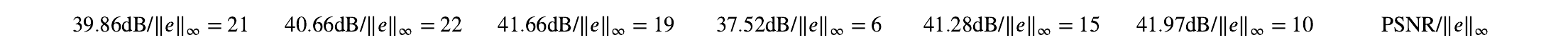}
\vskip -0.45cm
\caption{Visual comparisons of different methods. The second row presents close-up views.
Notice how the nuts (green arrows) on the hatch are erased by J2K, WebP, BPG, DnCNN but preserved by $\ell_\infty\mbox{-ED}^2$, and how the errors (red arrows) of CALIC are corrected by $\ell_\infty\mbox{-ED}^2$.}
\label{space}
\end{figure*}
\begin{figure*}[!ht]
\centering
\vskip -0.2cm
\includegraphics[width=\linewidth]{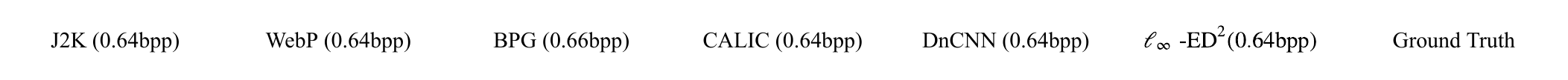}
\vskip -0.2cm
\includegraphics[width=\linewidth, height=0.14\linewidth]{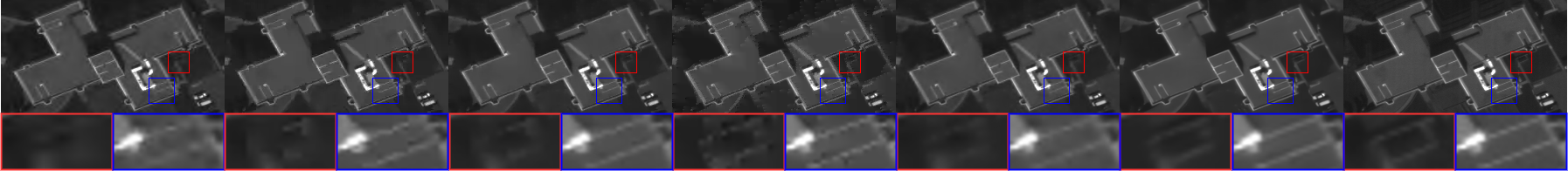}
\includegraphics[clip, trim=0.8cm 0cm 0.4cm 0cm, width=\linewidth]{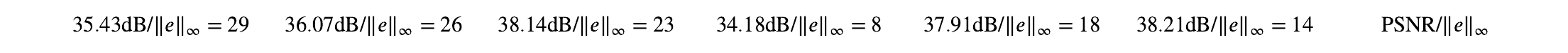}
\vskip -0.45cm
\caption{Visual comparisons of different methods.
Note the erasures and distortions of line structures by J2K, WebP, BPG, CALIC and DnCNN
in comparison with $\ell_\infty\mbox{-ED}^2$ in close-up views, and also the correction of
CALIC distortions by $\ell_\infty\mbox{-ED}^2$.}
\label{aerial01}
\end{figure*}
\begin{figure*}[!ht]
\centering
\vskip -0.2cm
\includegraphics[width=\linewidth]{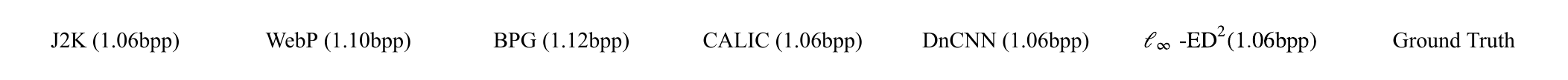}
\vskip -0.2cm
\includegraphics[width=\linewidth, height=0.14\linewidth]{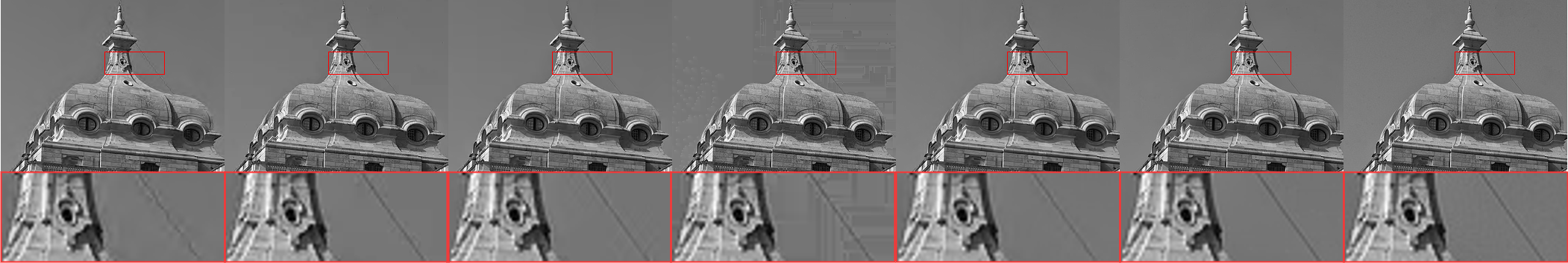}
\includegraphics[clip, trim=0.8cm 0cm 0.4cm 0cm, width=\linewidth]{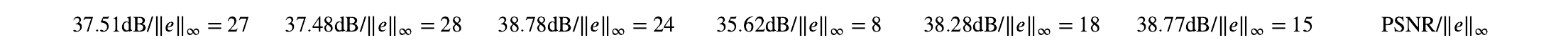}
\vskip -0.45cm
\caption{Visual comparisons of different methods.
Note the artifacts on and near the cable produced by J2K, WebP, BPG, CALIC and DnCNN
in comparison with $\ell_\infty\mbox{-ED}^2$ in close-up views, and also the correction of
CALIC's patterned errors in smooth background by $\ell_\infty\mbox{-ED}^2$.
}
\label{urban054}
\end{figure*}
\begin{figure*}[!ht]
\centering
\vskip -0.2cm
\includegraphics[width=\linewidth]{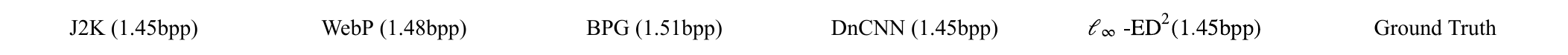}
\vskip -0.05cm
\includegraphics[width=\linewidth, height=0.15\linewidth]{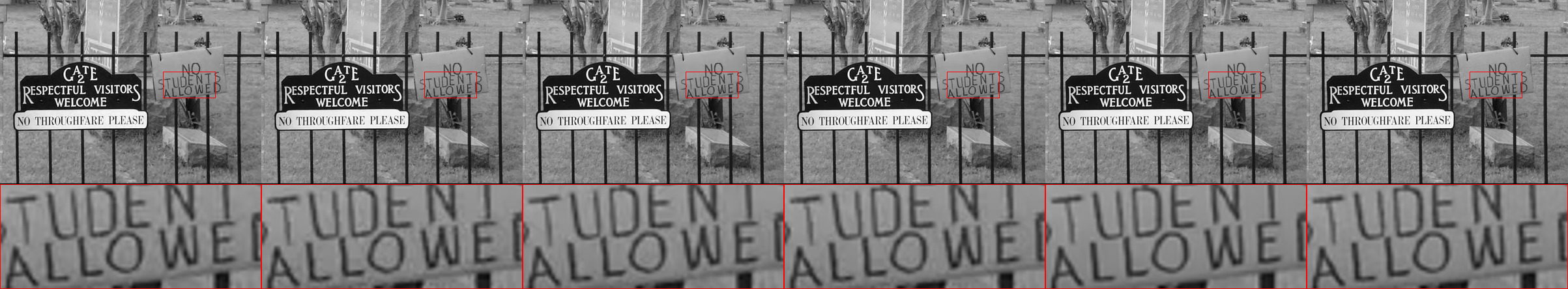}
\includegraphics[clip, trim=0.6cm 0cm 0.4cm 0cm, width=\linewidth]{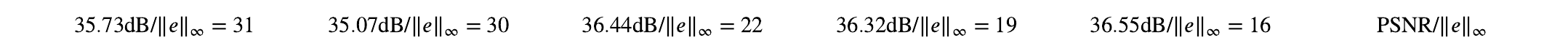}
\vskip -0.45cm
\caption{Visual comparisons of different methods.  In the close-up views, note
the ghost artifacts on the letters produced by J2K, WebP, BPG and DnCNN,
and clean letter reconstruction by $\ell_\infty\mbox{-ED}^2$.}
\label{cemetry}
\end{figure*}
\begin{figure*}[!ht]
\centering
\vskip -0.2cm
\includegraphics[width=\linewidth]{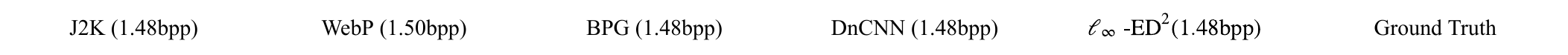}
\vskip -0.05cm
\includegraphics[width=\linewidth, height=0.15\linewidth]{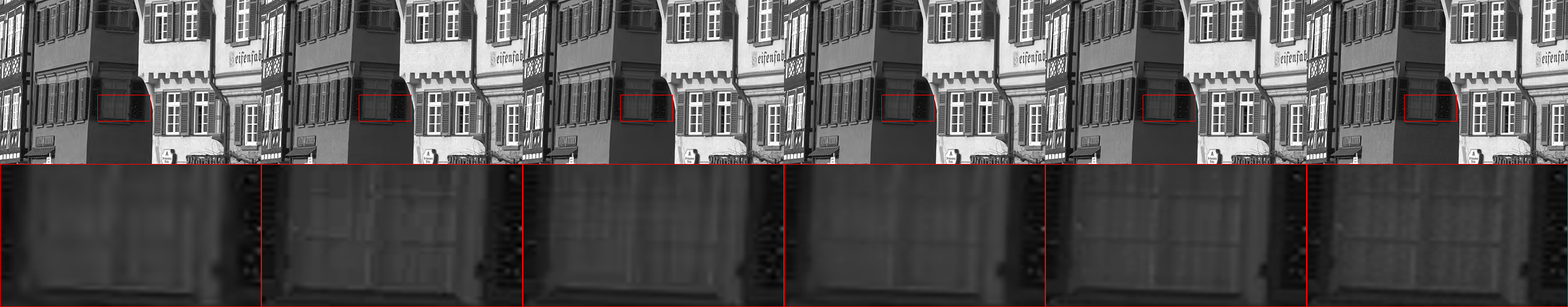}
\includegraphics[clip, trim=0.6cm 0cm 0.4cm 0cm, width=\linewidth]{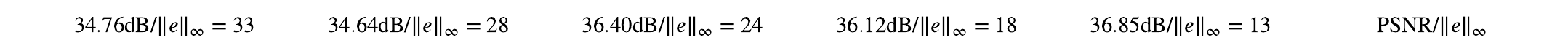}
\vskip -0.45cm
\caption{Visual comparisons of different methods. In the close-up views, note the vertical window ripplefold drapery pattern and the horizontal window bars which are distorted by the J2K, WebP, BPG and DnCNN, but faithfully reconstructed by $\ell_\infty\mbox{-ED}^2$.}
\label{buildings}
\end{figure*}

\subsection{Perceptual quality comparison}

The perceptual qualities of competing methods, given the same bit rate, are compared in Figs.~\ref{space} through \ref{buildings}.  Recall that our motivation is to devise a perceptually lossless image compression system that has a provable $\ell_\infty$ error bound.  Therefore, the comparisons are made near the critical bit rate for perceptually transparent reconstruction.  For each image the PSNR, $\ell_\infty$ error and the bit rate are stated.

The space shuttle test image (Fig.~\ref{space}) is a telling example for the necessity of using the $\ell_\infty$ loss in ultra-high fidelity image compression in scientific and medical applications.  The nuts on the shuttle hatch (see the green arrow in the ground truth image) are removed by WebP, BPG and DnCNN, which is an unacceptable semantic error.  At the same bit rate the near-lossless CALIC preserves these nuts but it generates some false dots (marked by red arrow).  Only the $\ell_\infty\mbox{-ED}^2$ compression system is flawless.

In the second example of aerial image (Fig.~\ref{aerial01}), in the red window, the faint rectangular structure in the original image is completely erased by J2K, and it becomes illegible in the outputs of WebP, BPG and DnCNN due to severe blurring, while the $\ell_\infty\mbox{-ED}^2$ system is able to recover the structure perfectly.
In the blue window of Fig.~\ref{aerial01}, J2K almost erases the original parallel lines; the other competing methods are not satisfactory either with severe blurring and jaggy artifacts.
Again, only the $\ell_\infty\mbox{-ED}^2$ compression system achieves perceptually lossless reconstruction.

In the example of the roof top image (Fig.~\ref{urban054}), all competing methods produce blurring, ghost and jaggy artifacts on the cable and roof contour in various degree, whereas
the $\ell_\infty\mbox{-ED}^2$ recovers the clean and sharp line and curves.  Note how the structured compression artifacts of CALIC in smooth background are removed by the CNN-based soft decoding methods.

In Fig.~\ref{cemetry}, the letters restored by the $\ell_\infty\mbox{-ED}^2$ compression system are much cleaner and sharper than other methods with a visually lossless quality with respect to the ground truth.

In Fig.~\ref{buildings}, after zooming in, one can see that the $\ell_\infty\mbox{-ED}^2$ compression system faithfully reconstructs the vertical window ripplefold drapery pattern and the horizontal window bars in the original image, but all other methods fail to do so, distorting the original patterns and structures noticeably.

%the two horizontal lines inside the window are almost erased by JPEG2000 and BPG. WebP preserves the lines but introduces some artifacts which severely degrade the visual quality. DnCNN produces a blurry image, failing to restore the two straight lines.  Only the proposed $\ell_\infty\mbox{-ED}^2$ system recovers the lines faithfully.

Thanks to the $\ell_\infty$ minmax criterion, the proposed $\ell_\infty$-SDNet soft decoding method can restore sharp edges and subtle features more accurately than the other methods, which explains its superior perceptual quality.

\subsection{Ablation study}
\begin{figure}[!t]
\centering
\vskip -0.2cm
\includegraphics[width=\linewidth]{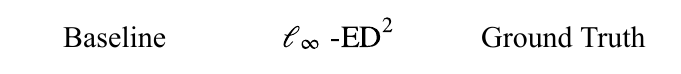}
\vskip -0.2cm
\includegraphics[width=\linewidth]{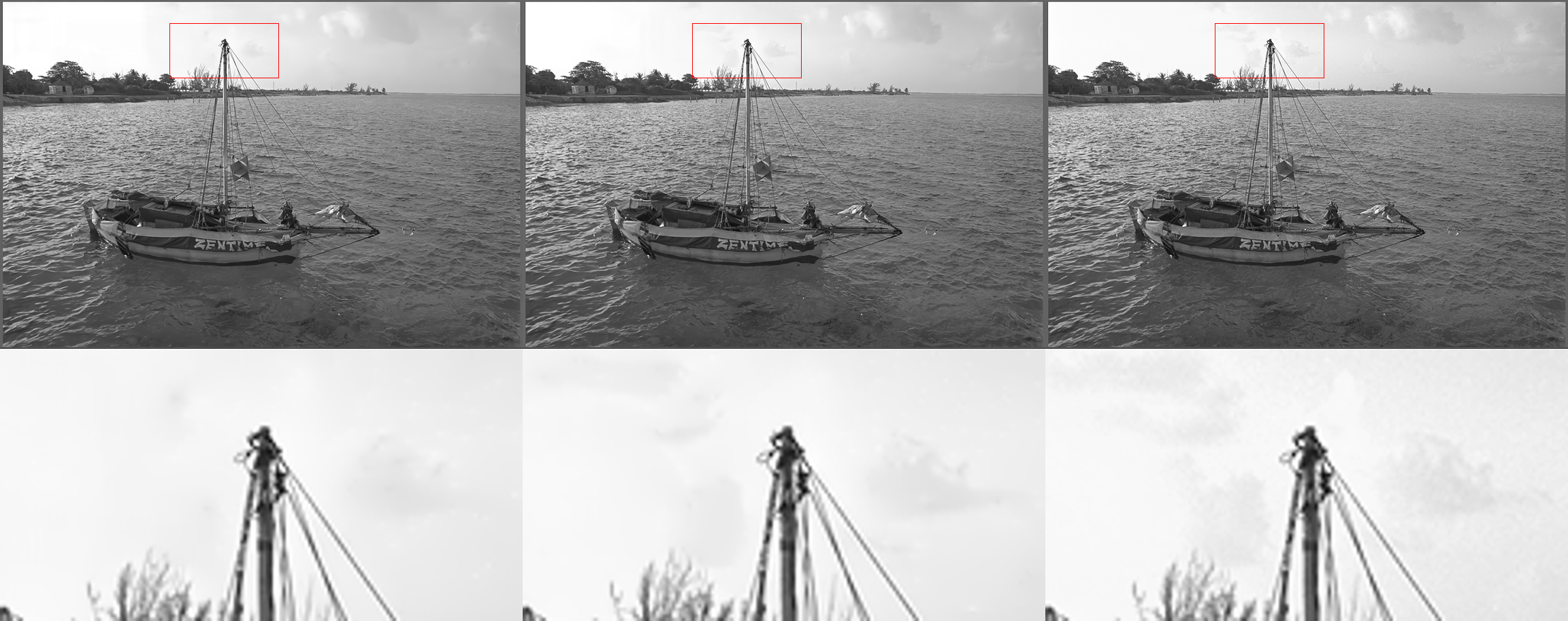}
\caption{Visual comparisons of different methods. Notice how the subtle clouds in the sky are almost erased by the baseline network, but well reconstructed by $\ell_\infty\mbox{-SDNet}$.}
\label{kodim06}
\end{figure}

\begin{figure}[!t]
\centering
\vskip -0.2cm
\includegraphics[width=\linewidth]{figure/cap_ablation.pdf}
\vskip -0.2cm
\includegraphics[width=\linewidth]{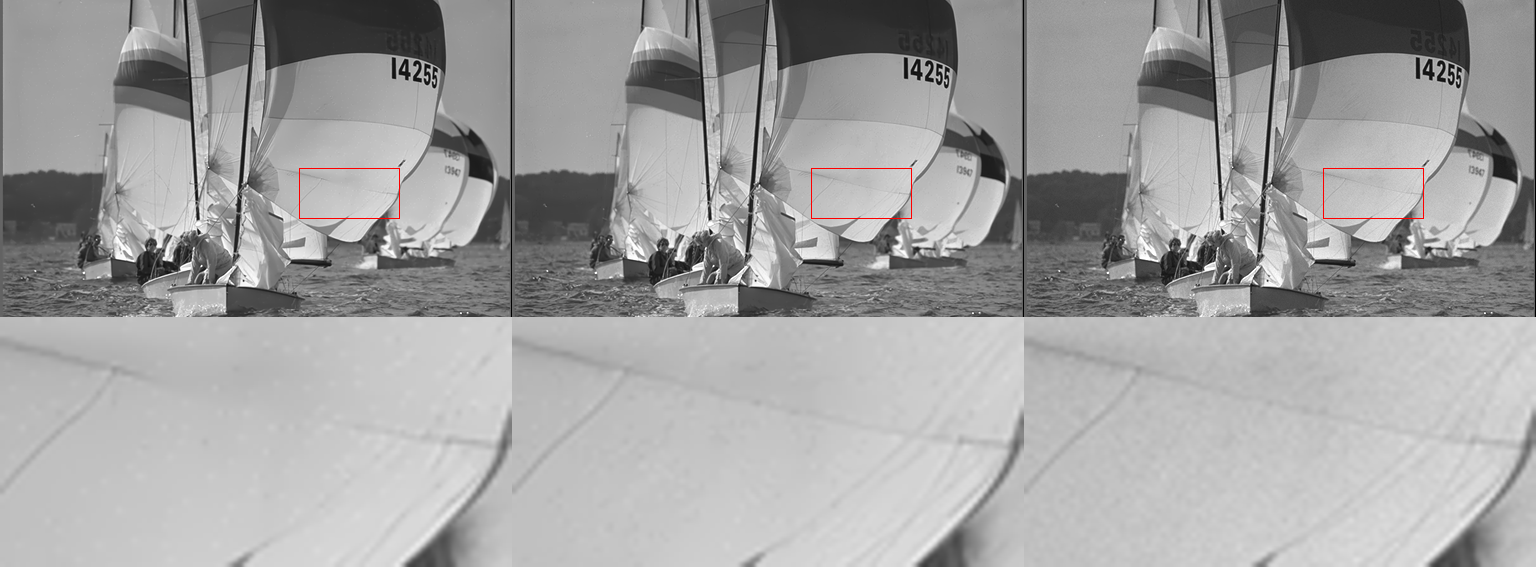}
\caption{Visual comparisons of different methods. Note the line on the sail which is broken by the baseline network, but recovered by the proposed $\ell_\infty\mbox{-SDNet}$ completely.}
\label{kodim09}
\end{figure}

\textbf{Quasi-$\ell_\infty$ loss and truncated activation}
In order to isolate the effects of imposing the $\ell_\infty$ loss, we build a baseline network of the same architecture as $\ell_\infty\mbox{-SDNet}$ but without the truncated activation module, and then train the baseline using only the MSE loss.  The performance results of the baseline network on Kodak dataset are tabulated in Table~\ref{tab_ablation} in comparison with those of the proposed $\ell_\infty\mbox{-SDNet}$. It can be seen that, after adding the quasi-$\ell_\infty$ loss and truncated activation, the network makes an appreciable performance gain in PSNR and $\ell_\infty$ metric.  We also provide the visual comparison results in Fig.~\ref{kodim06} and \ref{kodim09}. As shown in Fig.~\ref{kodim06}, the subtle clouds in the sky are almost erased by the baseline network, but are well reconstructed by $\ell_\infty\mbox{-SDNet}$. In Fig.~\ref{kodim09}, the line on the sail is broken in the image recovered by the baseline, but it is recovered by $\ell_\infty\mbox{-SDNet}$ completely, even though the signal is very weak.

\begin{table}[!t]
\caption{Performance results (PSNR/$\|e\|_\infty$) of the ablation studies on the Kodak dataset.}
\begin{center}
\renewcommand\arraystretch{1.3}
\begin{tabular}{c|ccccc}
\hline
BPP & Baseline & No dilation & Single-rate & $\ell_\infty\mbox{-ED}^2$ \\
\hline
\hline
2.61 & 50.20\ /\ 2.00 & 50.21\ /\ 2.00 & 50.24\ /\ 2.00 & 50.26\ /\ 2.00 \\
1.98 & 46.32\ /\ 4.00 & 46.29\ /\ 4.00 & 46.37\ /\ 4.00 & 46.42\ /\ 4.00 \\
1.60 & 44.06\ /\ 6.00 & 44.05\ /\ 5.95 & 44.12\ /\ 5.84 & 44.19\ /\ 5.83 \\
1.35 & 42.48\ /\ 8.00 & 42.41\ /\ 7.76 & 42.51\ /\ 7.30 & 42.61\ /\ 7.29 \\
1.19 & 41.24\ /\ 9.89 & 41.15\ /\ 9.54 & 41.29\ /\ 9.11 & 41.38\ /\ 9.04 \\
1.03 & 40.09\ /\ 11.74 & 40.01\ /\ 11.12 & 40.19\ /\ 10.88 & 40.29\ /\ 10.79 \\
0.91 & 39.12\ /\ 13.65 & 39.06\ /\ 12.96 & 39.25\ /\ 12.48 & 39.36\ /\ 12.38 \\
0.82 & 38.24\ /\ 15.68 & 38.21\ /\ 14.78 & 38.42\ /\ 14.25 & 38.55\ /\ 14.16 \\
\hline
\end{tabular}
\end{center}
\label{tab_ablation}
\end{table}

\textbf{Dilated convolution} We also conduct experiments to evaluate the effectiveness of the dilated convolution used in our network. We build a network by replacing the dilated convolution operation in $\ell_\infty\mbox{-SDNet}$ with the traditional convolution and train it with the same settings as in $\ell_\infty\mbox{-SDNet}$.  The performance results of using dilated convolution versus conventional convolution on Kodak dataset are compared in
Table~\ref{tab_ablation}.  As can be seen, the dilated convolution yields a small performance gain.

\textbf{Multi-rate training}
To evaluate the universality of the proposed multi-rate training, we train eight rate-specific $\ell_\infty\mbox{-SDNet}$s for eight different rates and compare these single-rate CNNs with the proposed multi-rate $\ell_\infty\mbox{-SDNet}$.  Comparison results are shown in Table~\ref{tab_ablation}.  It turns out, that the proposed multi-rate $\ell_\infty\mbox{-SDNet}$ is more robust than the single-rate counterpart.  The multi-rate $\ell_\infty\mbox{-SDNet}$ even outperforms the single-rate $\ell_\infty\mbox{-SDNet}$ on testing images that are compressed at the said single rate.

\section{Conclusion}
\label{sec:conclusion}
In this paper, we propose a novel image compression system of $\ell_\infty$-constrained encoding coupled with CNN-based soft decoding ($\ell_\infty\mbox{-ED}^2$).  The proposed $\ell_\infty \mbox{-ED}^2$ approach beats the best of existing lossy image compression methods (e.g., BPG, WebP, etc.) not only in $\ell_\infty$ but also in $\ell_2$ error metric and perceptual quality, for bit rates near the threshold of perceptually transparent reconstruction.

% or


% \appendices

% % you can choose not to have a title for an appendix
% % if you want by leaving the argument blank
% \section{}

% % use section* for acknowledgment
% \section*{Acknowledgment}


\ifCLASSOPTIONcaptionsoff
  \newpage
\fi



%\bibliographystyle{IEEEtran}



% \end{thebibliography}

\bibliographystyle{IEEEtran}
\bibliography{calic_ref}


% \begin{IEEEbiography}{Xi Zhang}
% \end{IEEEbiography}

% \begin{IEEEbiography}{Xiaolin Wu}
% \end{IEEEbiography}


%\vfill


\end{document}